\documentclass[a4paper,11pt]{article}
\pdfoutput=1

\usepackage{jheppub}
\usepackage{epsfig}
\usepackage[T1]{fontenc}
\usepackage{multirow}

\usepackage[utf8]{inputenc}

\usepackage{textcomp,color}
\usepackage{graphicx,subfigure}

\usepackage[normalem]{ulem}

\title{Multi-kink collisions in the $\phi^6$ model}

\author[a,1]{Aliakbar Moradi Marjaneh,\note{Corresponding author.}}
\author[b,c]{Vakhid A. Gani,}
\author[d]{Danial Saadatmand,}
\author[e,f]{\\ Sergey V. Dmitriev,}
\author[a]{Kurosh Javidan}

\affiliation[a]{Department of Physics, Quchan Branch, Islamic Azad University, Quchan, Iran}
\affiliation[b]{Department of Mathematics, National Research Nuclear University MEPhI (Moscow Engineering Physics Institute), 115409 Moscow, Russia}
\affiliation[c]{Theory Department, National Research Center Kurchatov Institute, Institute for Theoretical and Experimental Physics, 117218 Moscow, Russia}
\affiliation[d]{Department of Physics, University of Sistan and Baluchestan, Zahedan, Iran}
\affiliation[e]{Institute for Metals Superplasticity Problems RAS, 450001 Ufa, Russia}
\affiliation[f]{National Research Tomsk State University, Tomsk 634050, Russia}

\emailAdd{moradimarjaneh@gmail.com}
\emailAdd{vagani@mephi.ru}
\emailAdd{saadatmand.d@gmail.com}
\emailAdd{dmitriev.sergey.v@gmail.com}
\emailAdd{javidan@um.ac.ir}

\abstract{We study simultaneous collisions of two, three, and four kinks and antikinks of the $\phi^6$ model at the same spatial point. Unlike the $\phi^4$ kinks, the $\phi^6$ kinks are asymmetric and this enriches the variety of the collision scenarios. In our numerical simulations we observe both reflection and bound state formation depending on the number of kinks and on their spatial ordering in the initial configuration. We also analyze the extreme values of the energy densities and the field gradient observed during the collisions. Our results suggest that very high energy densities can be produced in multi-kink collisions in a controllable manner. Appearance of high energy density spots in multi-kink collisions can be important in various physical applications of the Klein-Gordon model.}

\makeatletter
\def\@fpheader{\relax}
\makeatother

\begin{document}

\maketitle

\flushbottom

\section{Introduction}\label{sec:Introduction}

Nonlinear field models are of great interest in various areas of modern physics. Such models can have topological defects (or topological solitons) --- solutions that are homotopically distinct from the vacuum \cite{vilenkin01,manton01,aek01}. Such defects (domain walls, strings, vortices, kinks) arise, for example, in high energy physics, cosmology, condensed matter, and so on. We notice also the impressive progress in various scenarios with embedded topological defects, e.g., a Q-lump on a domain wall, or a skyrmion on a domain wall~\cite{nitta1,nitta2,kur01,blyankinshtein,kur02,nitta3,nitta4,jennings,nitta5,GaLiRa,GaLiRaconf}.

Special attention is paid in this context to (1+1)-dimensional models. Many realistic models in (3+1) and (2+1) dimensions can be reduced to the effective (1+1)-dimensional dynamics. For example a two- or three-dimensional domain wall in the direction orthogonal to it can be viewed as a topological soliton (kink) interpolating two different vacua of the model, which are separated by the wall in the two- or three-dimensional world. Besides that, (1+1)-dimensional models can be used as a simplified setup for studying general properties of nonlinear field models \cite{GaLiRa,GaLiRaconf,nitta6,GaKiRu,lensky,GaKsKu01,GaKsKu02}.

Topological solitons (kinks) in (1+1)-dimensional field models have been actively studied recently \cite{Kudryavtsev1975,Anninos1991,Goodman2007,Campbell1983,Peyrard1983,Campbell1986,dorey,GaKuLi,GaKuPRE,oliveira01,oliveira02,krusch01,saad01,saad02,saad03,rad1,rad2,saad.arXiv.2016.08,Deformed,Ahlqvist:2014uha,Mohammadi}. In particular, the kink-(anti)kink scattering and the interactions of kinks with impurities are of growing interest. A wide variety of phenomena emerges in these systems, e.g., escape windows and quasi-resonances in kink-(antikink) collisions \cite{Kudryavtsev1975,Anninos1991,Goodman2007,Campbell1983,Peyrard1983,Campbell1986,dorey,GaKuLi,GaKuPRE,oliveira01,oliveira02}, resonant interactions of kinks with wells, barriers and impurities \cite{krusch01,saad01,saad02,saad03}, non-radiative energy exchange in multi-soliton collisions \cite{rad1,rad2,saad.arXiv.2016.08}. It is interesting that the presence of a kink's internal modes does not guarantee the appearance of resonance windows, as it has been recently shown for the deformed $\phi^4$ model \cite{Deformed}.

The interactions of kinks (and antikinks) are studied using different methods, in particular, quasi-exact methods such as the numerical solution of the equations of motion, which are partial differential equations, and approximate methods such as the collective coordinate approximation \cite{GaKuLi,weigel01,weigel02,baron01,javidan,christov01,GaKu} and the Manton's method \cite{manton_npb,kks04,Radomskiy}. The simple collective coordinate approximation describes the dynamics of the kink-(anti)kink system in terms of the time dependence of the distance between the kinks, while the Manton's method allows one to estimate the interaction between the kink and the (anti)kink at large distances by the use of the kinks' asymptotics.

The dynamic properties of the kink-antikink collisions have been extensively investigated in integrable and non-integrable models. It was shown that the energy loss due to the radiation during the collision is small in integrable models. In contrast to that, in non-integrable models the radiation effects become important. The amount of radiation is a complicated function of the initial velocity, and depending on it the result of a kink-antikink collision can be very different: the solitons can form an oscillating bound state, or they can bounce back and reflect from each other. The collision process in non-integrable models is chaotic, in the sense that at some values of the initial velocities the kinks scatter off each other, while at other initial velocities they can annihilate. This behavior is a consequence of resonances between the oscillations of the kink's pairs and the excitations of their vibrational modes. There is no exhaustive theoretical model for describing all the features of collisions of solitons in non-integrable models, and the best way to investigate the properties of such systems is numerical simulation.

In our previous publications \cite{Aliakbar,saad.prd.2015} we have studied multi-kink collisions in the sine-Gordon and $\phi^4$ models. We have shown that the maximal values of the total energy density can be achieved if all $N$ kinks/antikinks collide at the same point. This happens when the kinks and antikinks approach the collision point in an alternating order (i.e.\ no two adjacent solitons are of the same type). When arranged in this way, the solitons attract each other and their cores can merge producing high energy density spots. The effect of the kink's internal mode on the maximal total energy density was studied for the $\phi^4$ model \cite{Aliakbar}. It has been shown that the kink's internal mode can increase or decrease the value of the energy density that can be produced at the collision point.

In this paper we study the collisions of $N\leq 4$ kinks of the $\phi^6$ model numerically. The (1+1)-dimensional $\phi^6$ model is well-known in the literature \cite{lohe,dorey,GaKuLi,weigel01,weigel02}, but the study of simultaneous multi-kink collisions in this model is carried out for the first time. A simultaneous collision of several kinks in a small region can produce a very large energy density in this region. Such regions could be of great interest in studying various physical systems described by (1+1)-dimensional field-theoretical models.

Our paper is organized as follows. In section \ref{sec:Model} we briefly describe the (1+1)-dimensional $\phi^6$ model and its topologically non-trivial solutions --- kinks and antikinks. In section \ref{sec:Results} we describe our method and present the results of the numerical study of the collisions of $N=2$ (subsection \ref{Sec:N2}), $N=3$ (subsection \ref{Sec:N3}), and $N=4$ (subsection \ref{Sec:N4}) kinks at the same point. In section \ref{sec:Conclusion} we give the conclusion and an outlook.

\section{The model}\label{sec:Model}

The $\phi^6$ model in $(1+1)$-dimensional space-time is described by the Lagrangian density
\begin{equation}\label{eq:Largangian}
\mathcal{L} = \frac{1}{2}\left(\frac{\partial\phi}{\partial t}\right)^2 - \frac{1}{2}\left(\frac{\partial\phi}{\partial x}\right)^2 - V(\phi),
\end{equation}
where $\phi(x,t)$ is a real scalar field. The potential $V(\phi)$, which defines the self-interaction of the field, has the form
\begin{equation}\label{eq:potential}
V(\phi)=\frac{1}{2}\phi^2(1-\phi^2)^2.
\end{equation}
\begin{figure}[h]
\centering
\includegraphics[width=0.5\textwidth]{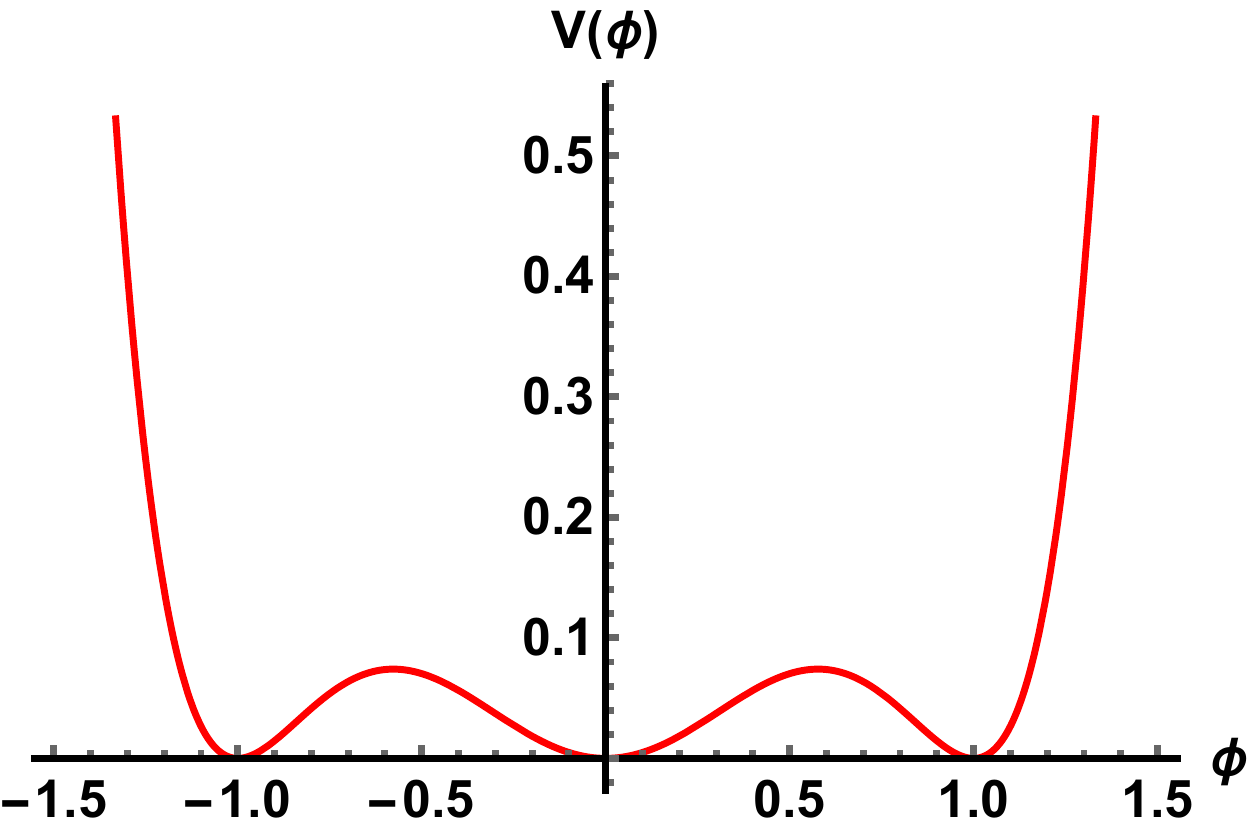} 
\caption{The potential \eqref{eq:potential} of the $\phi^6$ model.} 
\label{fig:Phi6VvsPhi}
\end{figure}
The energy functional corresponding to the Lagrangian \eqref{eq:Largangian} is
\begin{equation}\label{eq:energy}
E[\phi]=\int_{-\infty}^{+\infty}\left[\frac{1}{2}\left(\frac{\partial\phi}{\partial t}\right)^2 + \frac{1}{2}\left(\frac{\partial\phi}{\partial x}\right)^2 + V(\phi)\right]dx.
\end{equation}
The Lagrangian \eqref{eq:Largangian} yields the equation of motion for the field $\phi(x,t)$:
\begin{equation}\label{eq:eom}
\frac{\partial^2\phi}{\partial t^2} - \frac{\partial^2\phi}{\partial x^2} + \frac{dV}{d\phi}=0.
\end{equation}
The potential \eqref{eq:potential} is non-negative function, and it has three degenerate minima (vacua of the model): $\bar{\phi}_1=-1$, $\bar{\phi}_2=0$, and $\bar{\phi}_3=1$, with $V(\bar{\phi}_1)=V(\bar{\phi}_2)=V(\bar{\phi}_3)=0$, see figure \ref{fig:Phi6VvsPhi}. Therefore the model has topological soliton solutions (kinks) --- static field configurations $\phi_\mathrm{K}(x)$ interpolating between neighboring vacua.

We use the following notation: a kink $\phi_\mathrm{K}(x)$ is said to belong to the topological sector $(\bar{\phi}_i,\bar{\phi}_j)$ if $\lim\limits_{x\to-\infty} \phi_\mathrm{K}(x)=\bar{\phi}_i$ and $\lim\limits_{x\to+\infty} \phi_\mathrm{K}(x)=\bar{\phi}_j$. We also denote this kink by using $\phi_{(\bar{\phi}_i,\bar{\phi}_j)}(x)$ instead of $\phi_\mathrm{K}(x)$.

\begin{figure}[h]
\centering
\includegraphics[width=0.5\textwidth]{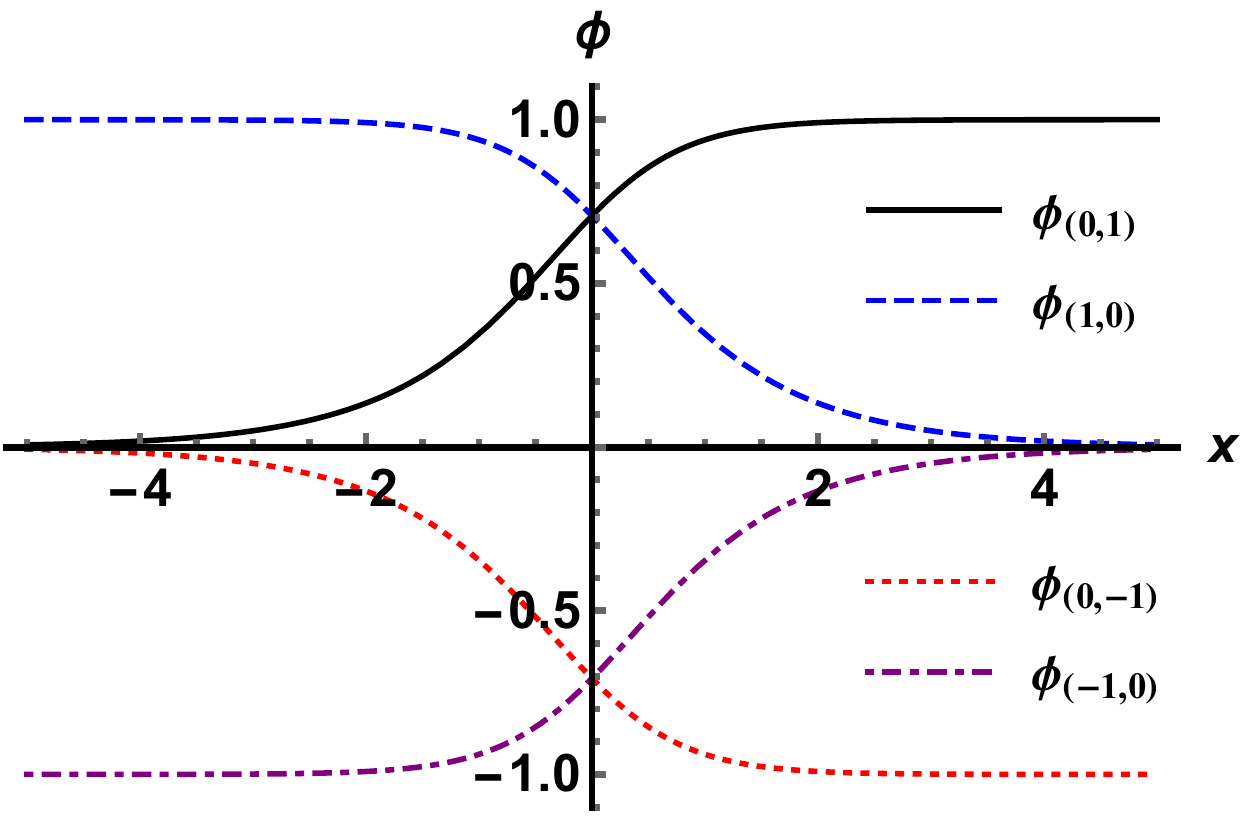} 
\caption{Kinks and antikinks of the $\phi^6$ model.} 
\label{fig:Phi6kinks}
\end{figure}

The kinks of the $\phi^6$ model can be easily found analytically by solving eq.~\eqref{eq:eom}. In the static case $\displaystyle\frac{\partial\phi}{\partial t}=0$, and we obtain
\begin{equation}
\frac{d^2\phi}{dx^2}=\frac{dV}{d\phi}.
\end{equation}
This equation can be reduced to the first order ordinary differential equation
\begin{equation}
\frac{d\phi}{dx}=\pm\sqrt{2V(\phi)}.
\end{equation}
Because the potential \eqref{eq:potential} has three minima, there are two kinks and two antikinks in the model, see figure \ref{fig:Phi6kinks}.
The kinks belong to the topological sectors $(-1,0)$ and $(0,1)$, while the antikinks belong to the sectors $(0,-1)$ and $(1,0)$. Below in some cases we use the term ``kink'' instead of the term ``antikink'' for brevity.

For future convenience, we write out here all static kinks and antikinks of the $\phi^6$ model:
\begin{equation}\label{eq:kinks1}
\phi_{(0,1)}(x) =\sqrt{\frac{1+\tanh x}{2}},
\quad
\phi_{(1,0)}(x) = \sqrt{\frac{1-\tanh x}{2}},
\end{equation}
\begin{equation}\label{eq:kinks2}
\phi_{(-1,0)}(x) = -\sqrt{\frac{1-\tanh x}{2}},
\quad
\phi_{(0,-1)}(x) = -\sqrt{\frac{1+\tanh x}{2}}.
\end{equation}
Notice that the kinks of the $\phi^6$ model are asymmetric with respect to the spatial reflection. Consider, e.g., the kink $\phi_{(0,1)}(x)$. At $|x|\gg 1$ we have the following asymptotics:
\begin{equation}\label{eq:asymptotics1}
\phi_{(0,1)}(x)\sim e^x, \quad x\to-\infty,
\end{equation}
\begin{equation}\label{eq:asymptotics2}
\phi_{(0,1)}(x)\sim 1-\frac{1}{2}e^{-2x}, \quad x\to+\infty.
\end{equation}

The mass of each (anti)kink is $M_\mathrm{K}=\displaystyle\frac{1}{4}$, as can be obtainted by substituting eqs.~\eqref{eq:kinks1} or \eqref{eq:kinks2} into the energy functional \eqref{eq:energy}. A moving kink or antikink can be obtained from eqs.~\eqref{eq:kinks1} and \eqref{eq:kinks2} by the Lorentz boost. Assume that the kinks \eqref{eq:kinks1}, \eqref{eq:kinks2} are moving along the $x$-axis with the velocity $v$. Then for such moving kinks we have
\begin{equation}
\phi_{(0,1)}(x,t) = \phi_{(0,1)}(\gamma(x-vt))\quad \mbox{and so forth},
\end{equation}
where $\gamma = 1/\sqrt{1-v^2}$ is the Lorentz factor.

The total energy \eqref{eq:energy} can be split into three parts: the kinetic energy $K$, the gradient energy $U$, and the potential energy $P$,
\begin{equation}\label{Energy}
E = K + U + P.
\end{equation}
To do that, the integrand in \eqref{eq:energy}, i.e.\ the total energy density $\varepsilon(x,t)$, is written as
\begin{equation}\label{Edensity}
\varepsilon(x,t) = k(x,t) + u(x,t) + p(x,t),
\end{equation}
where
\begin{equation}
k(x,t) = \frac{1}{2}\left(\frac{\partial\phi}{\partial t}\right)^2,
\quad
u(x,t) = \frac{1}{2}\left(\frac{\partial\phi}{\partial x}\right)^2,
\quad
p(x,t) = \frac{1}{2}\phi^2(1-\phi^2)^2
\end{equation}
are the kinetic energy density, the gradient energy density, and the potential energy density, respectively. The gradient energy density, in turn, can be expressed as
\begin{equation}
u(x,t) = \frac{1}{2}e^2(x,t),
\end{equation}
where
\begin{eqnarray}\label{Strain}
e(x,t)=\frac{\partial\phi}{\partial x}
\end{eqnarray}
is the field gradient, which can be positive or negative corresponding to ``stretching'' or ``compression''.

For example, in the case of one moving kink $\phi_{(0,1)}(x,t)$ we have:
\begin{equation}
p(x,t) = \frac{1}{16}\:\frac{1-\tanh[\gamma(x-vt)]}{\cosh^2[\gamma(x-vt)]} = \frac{1-v^2}{v^2}\:k(x,t) = (1-v^2)\:u(x,t).
\end{equation}
The total energy density is
\begin{equation}
\varepsilon(x,t) = \frac{1}{8}\:\frac{1}{1-v^2}\:\frac{1-\tanh[\gamma(x-vt)]}{\cosh^2[\gamma(x-vt)]}.
\end{equation}
Integrating this expression with respect to $x$ over the interval from $-\infty$ to $+\infty$, we obtain the total energy of the moving kink:
\begin{equation}
E_\mathrm{K} = \int_{-\infty}^{+\infty}\varepsilon(x,t)\:dx = \frac{M_\mathrm{K}}{\sqrt{1-v^2}},
\end{equation}
where $M_\mathrm{K}=\displaystyle\frac{1}{4}$ is the mass of kink, i.e.\ the energy of the static $\phi^6$ kink.

Collisions of the kinks of the $(1+1)$-dimensional $\phi^6$ model can be investigated numerically. In the next section,  we present the results of our numerical simulations of collisions of two, three, and four kinks at the same point. Our goal is to find the maximal (over the spatial coordinate $x$ and the temporal coordinate $t$) values of the energy densities: kinetic, gradient, potential, and total. We also find the maximal values of the field gradient for each collision. Note that for one moving kink these extreme values can be obtained analytically:
\begin{equation}\label{eq:energy1}
p_\mathrm{max}^{(1)} = \frac{2}{27},
\quad
k_\mathrm{max}^{(1)} = \frac{2}{27}\frac{v^2}{1-v^2},
\quad
u_\mathrm{max}^{(1)} = \frac{2}{27}\frac{1}{1-v^2},
\quad
\varepsilon_\mathrm{max}^{(1)} = \frac{4}{27}\frac{1}{1-v^2},
\end{equation}
and
\begin{equation}\label{eq:strain1}
e_\mathrm{max}^{(1)} = \frac{2}{3\sqrt{3}}\frac{1}{\sqrt{1-v^2}}.
\end{equation}

\section{Numerical results}
\label{sec:Results}

We study the collisions of several $\phi^6$ kinks and antikinks at the same point (in a small region, to be more accurate). Because of the absence of analytic multisolitonic solutions for the $\phi^6$ model, we use initial conditions in the form of superposition of several kinks and antikinks, which are moving towards the collision point. For every event we adjust the initial positions and initial velocities of kinks to ensure their collision at one point.

The initial distances between the kinks in our simulations are quite large. Therefore the overlap of the kinks is exponentially small, i.e.\ the initial configurations are solutions of eq.~\eqref{eq:eom} with exponential accuracy.

For the numerical study of the evolution of the initial configurations we use the discretized version of the equation of motion \eqref{eq:eom}:
\begin{eqnarray}\label{Phi4discrete}
\frac{d^2\phi_n}{dt^2} - \frac{1}{h^2}(\phi_{n-1} -2\phi_{n}+\phi_{n+1}) 
&+\displaystyle\frac{1}{12h^2}(\phi_{n-2}-4\phi_{n-1} +6\phi_{n}-4\phi_{n+1}+\phi_{n+2})\nonumber\\
&+\phi_n(1-\phi_n^2)(1-3\phi_n^2)  = 0,
\end{eqnarray}
where $h$ is the lattice spacing, $n=0,\pm1,\pm2,...$, and $\phi_n(t)=\phi(nh,t)$. In order to minimize the finite-spacing artifacts, the term $\phi_{xx}$ in eq.~\eqref{Phi4discrete} is discretized with the accuracy $O(h^4)$ \cite{saad.prd.2015}. The equations of motion~\eqref{Phi4discrete} are integrated with respect to time using an explicit scheme with the time step $\tau$ (the value used in the calculations is $\tau=0.005$) and the accuracy $O(\tau^4)$. We use the St\"ormer method for the integration of eq.~\eqref{Phi4discrete}. In order to be sure that the maximal values of the energy density and the field gradient converge, we perform numerical simulations with different lattice spacings: $h = 0.1$, $h = 0.05$.

In the numerical simulations presented in this section we use $5000$ points for the spatial grid, corresponding to the $x$ range from $-250$ to $250$ for $h=0.1$ and from $-125$ to $125$ for $h=0.05$. Hence the spatial boundaries are far away enough and cannot affect the numerical results. In addition, we also use the absorbing boundary conditions in order to prevent the small amplitude radiation reflected from the boundaries. Below we report the results of the maximal energy densities and extreme values of the field gradient in the collisions of $N$ slow-moving kinks and antikinks for $1<N \leq 4$ [for the case $N=1$ these values can be found analytically, see eqs.~\eqref{eq:energy1} and \eqref{eq:strain1}].

Notice that in some figures we cut off high peaks in order to show the whole space-time picture of the collision better.

\subsection{Collision of two kinks}
\label{Sec:N2}

\subsubsection{The configuration (0,1,0)}

Now we study the collision of the kink $(0,1)$ and the antikink $(1,0)$, the initial configuration denoted as $(0,1,0)$, or $K\bar{K}$. According to this, the initial condition is taken as the kink $\phi_{(0,1)}(x-x_1,t)$ and antikink $\phi_{(1,0)}(x-x_2,t)$, placed at $x_1=-10$ and $x_2=10$. The initial velocities are $v_1=0.1$ and $v_2=-0.1$. As already mentioned, there is no exact two-soliton solution in the $\phi^6$ model, and we use the following initial configuration:
\begin{equation}\label{Kink010}
\phi_{(0,1,0)}(x,t) = \phi_{(0,1)}(x-x_1,t)+\phi_{(1,0)}(x-x_2,t)-1.
\end{equation}
At $|x_1-x_2|\gg 1$, this is a solution of the equation of motion up to the exponentially small overlap of the kink and the antikink.

In the collisions of the kinks $(0,1)$ and $(1,0)$, there is a critical value of the initial velocity $v_\mathrm{cr}\approx 0.289$ that separates two different regimes of the collision process. If the initial velocity is less than $v_\mathrm{cr}$ then the kink and antikink become trapped after the collision, forming a bound state (a bion). At initial velocities larger than $v_\mathrm{cr}$ the kink and the antikink escape to infinity after the collision. For further details see, e.g., refs.~\cite{dorey,GaKuLi} and references therein.

In our numerical simulation the initial velocity is $v_1=-v_2=0.1$, which is less than $ v_\mathrm{cr}$. This means that a bion should be formed after the collision. The bion stays near the collision point and emits energy in the form of small-amptlidude waves. We thus observe the ``reaction'' $K\bar{K}\to b$, where $b$ stands for the bion.

The numerical results for the configuration $(0,1,0)$ are presented in figure \ref{fig:010}. In figure \ref{fig:phi_010} we show the dependence $\phi(x,t)$, which demonstrates the main features of the collision process. Figure \ref{fig:total_energy_010} demonstrates the space-time dependence of the total energy density $\varepsilon(x,t)$. From this figure we see that the two peaks of the energy density are moving towards each other, collide, and form a bound state, which decays slowly emitting small waves. In figures \ref{fig:kinetic_energy_010}--\ref{fig:gradient_energy_010} we give the space-time pictures of the kinetic, potential, and gradient energy densities.

From figure \ref{fig:kinetic_energy_010} we see that the kinetic energy density of the moving kinks before the collision is small compared with the amplitude values of the subsequent kinetic energy oscillations in the bound state. The gradient energy density behaves differently, see figure \ref{fig:gradient_energy_010}:
\begin{figure}[h!]
\begin{center}
  \centering
  \subfigure[two kinks collision in the sector $(0,1,0)$]{\includegraphics[width=0.49\textwidth]{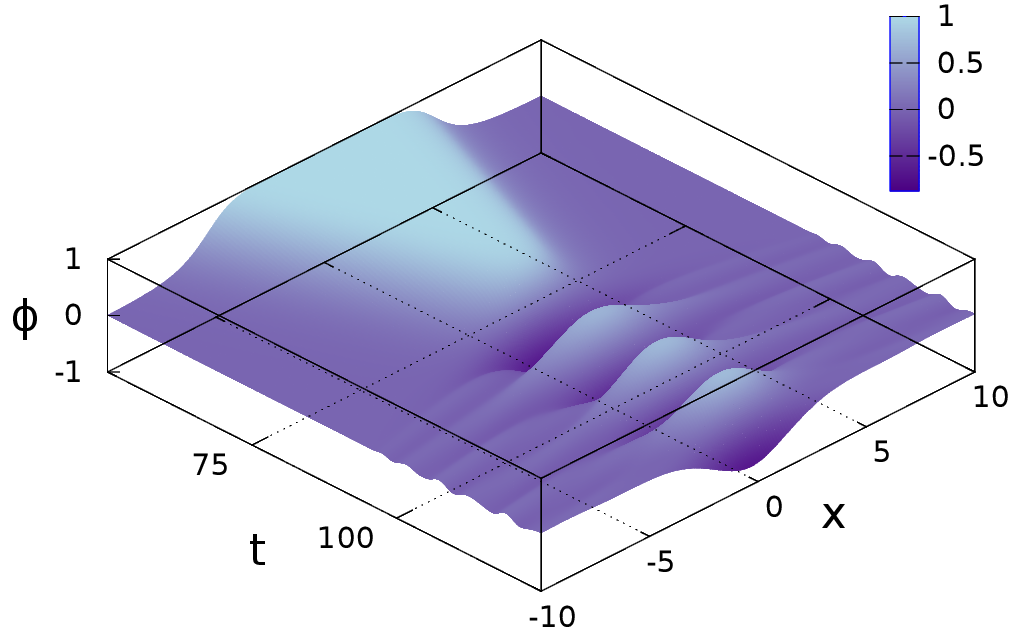}\label{fig:phi_010}}
  \subfigure[total energy density]{\includegraphics[width=0.49\textwidth]{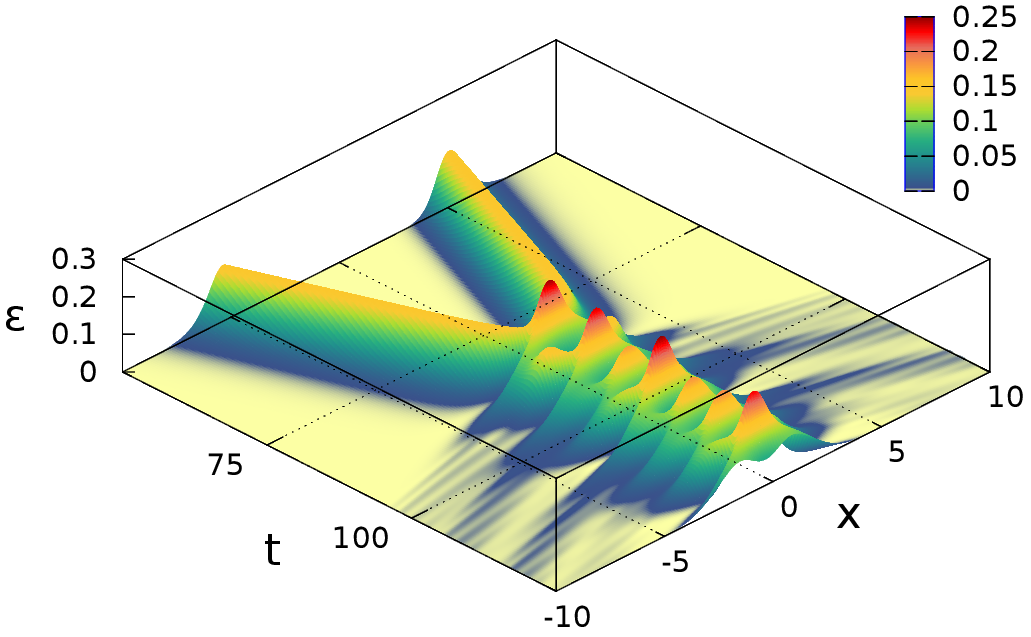}\label{fig:total_energy_010}}
  \\
  \subfigure[kinetic energy density]{\includegraphics[width=0.49\textwidth]{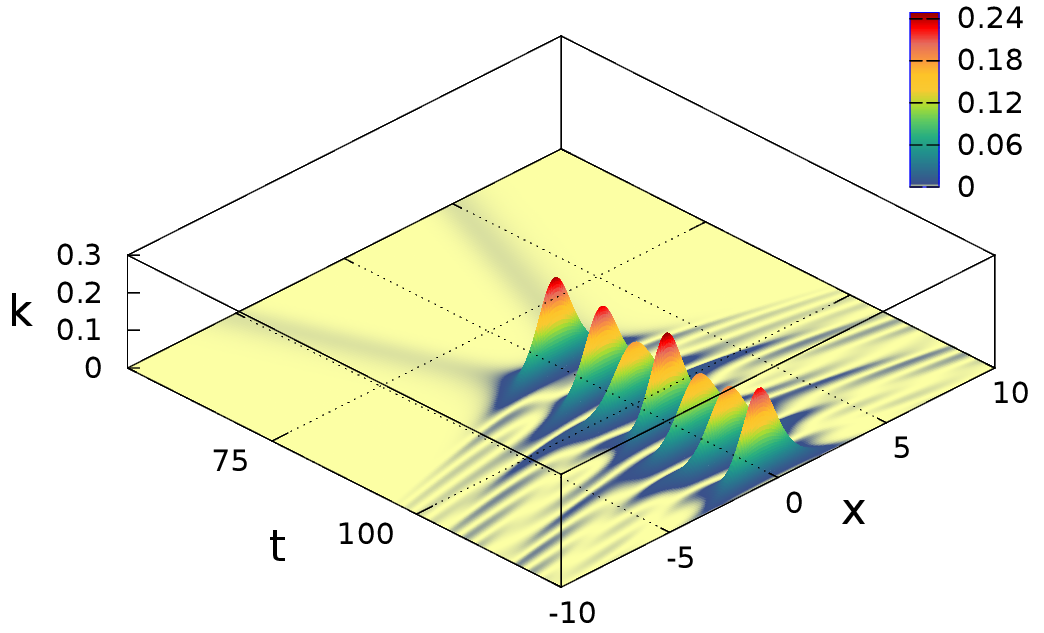}\label{fig:kinetic_energy_010}}
  \subfigure[potential energy density]{\includegraphics[width=0.49\textwidth]{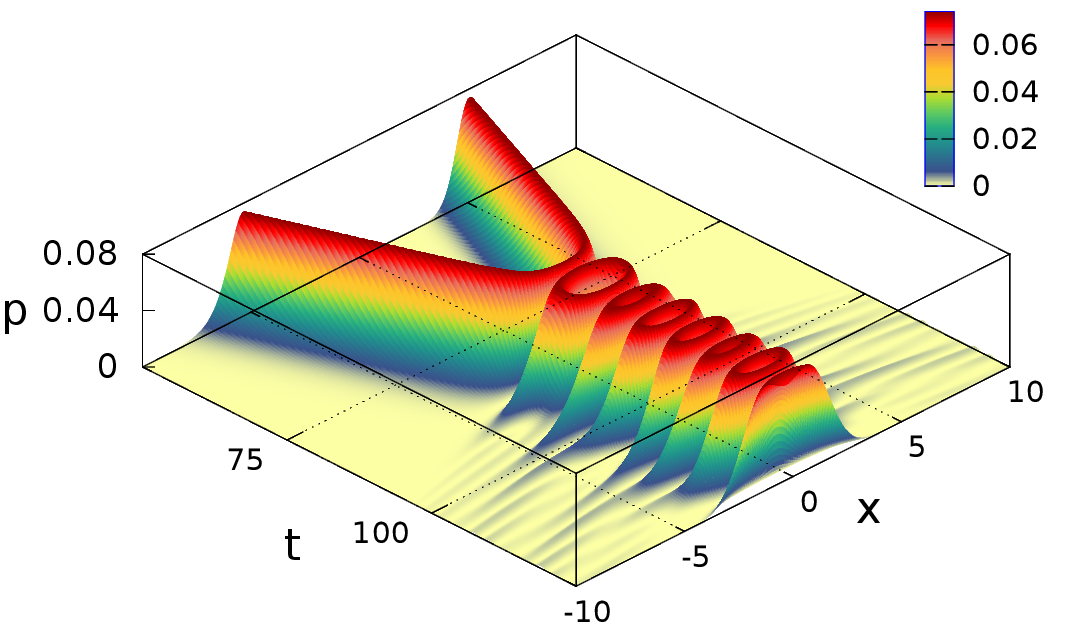}\label{fig:potential_energy_010}}
  \\
  \subfigure[gradient energy density]{\includegraphics[width=0.49\textwidth]{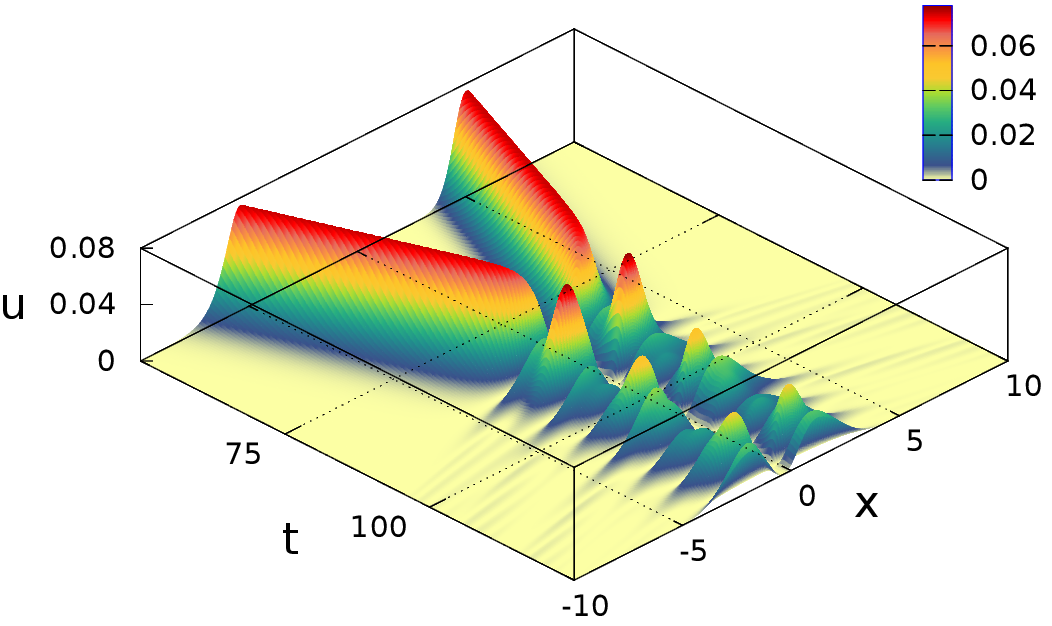}\label{fig:gradient_energy_010}}
  \subfigure[field gradient]{\includegraphics[width=0.49\textwidth]{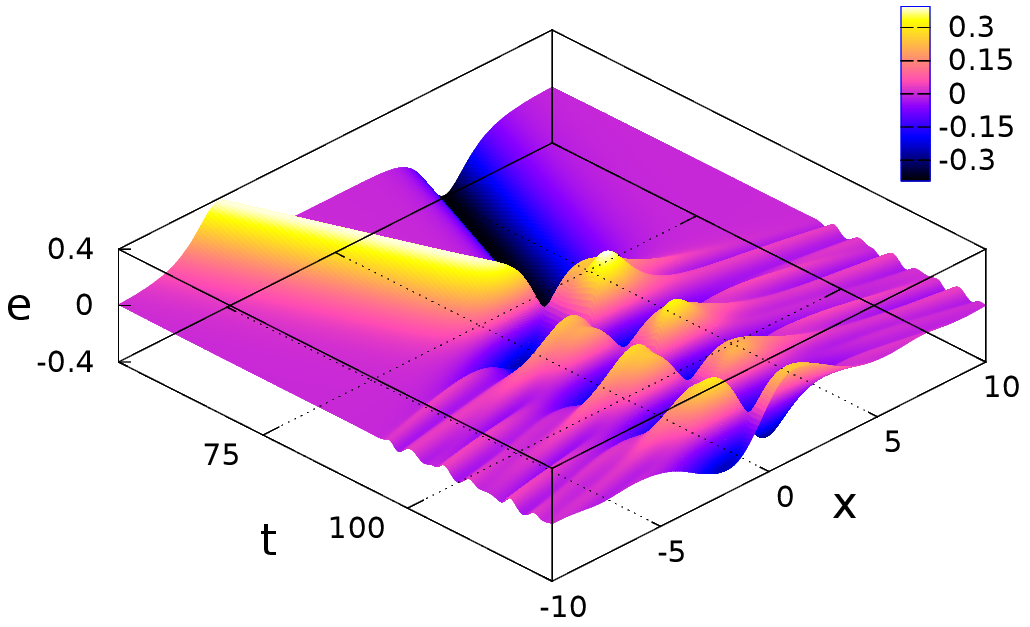}\label{fig:elastic_strain_010}}
  \caption{Space-time picture of collision of two kinks in the case of initial configuration $(0,1,0)$.}
  \label{fig:010}
\end{center}
\end{figure}
it decreases noticeably after the formation of the bound state. In figure \ref{fig:potential_energy_010} we show the potential energy density. It can be seen that it also oscillates but the amplitude falls off more slowly than in the previous case.

The space-time picture of the field gradient is shown in figure \ref{fig:elastic_strain_010}. In terms of the elastic strain, the kink and the antikink resemble a wave of compression and a wave of decompression, travelling towards each other. A localized oscillating structure is formed after the collision, see figure \ref{fig:elastic_strain_010}.

From the numerical analysis we obtain the following maximal values of the energy densities:
\begin{equation}
k_\mathrm{max}^{(2)} \approx 0.25,
\quad
u_\mathrm{max}^{(2)} \approx 0.075,
\quad
p_\mathrm{max}^{(2)} \approx 0.075,
\quad
\varepsilon_\mathrm{max}^{(2)} \approx 0.25.
\end{equation}
For the field gradient we found
\begin{equation}
e_\mathrm{min}^{(2)} \approx -0.4,
\quad
e_\mathrm{max}^{(2)} \approx 0.4.
\end{equation}

We have also performed the numerical simulation of the collision of the kinks $(0,-1)$ and $(-1,0)$ with the same initial velocities and initial positions, observing the same maximal values of the energy densities and the field gradient.

\subsubsection{The configuration (1,0,1)}

In this case, we use the initial configuration
\begin{equation}\label{Kink101}
\phi_{(1,0,1)}(x,t) = \phi_{(1,0)}(x-x_1,t)+\phi_{(0,1)}(x-x_2,t)
\end{equation}
in order to study the collisions of the kinks $(1,0)$ and $(0,1)$. We use the same values of $x_1$, $x_2$, $v_1$, and $v_2$ as in the previous subsection. The critical velocity value in this case is smaller than in the previous configuration, namely, $v_\mathrm{cr}\approx 0.045$~\cite{dorey,GaKuLi}. The initial velocity of the kinks in our numerical experiment is more than the critical value. This means that the kink and the antikink collide and escape from each other after the collision, i.e.\ we observe the ``reaction'' $\bar{K}K\to\bar{K}K$.

\begin{figure}[h!]
  \centering
  \subfigure[two kinks collision in the sector $(1,0,1)$]{\includegraphics[width=0.49\textwidth]{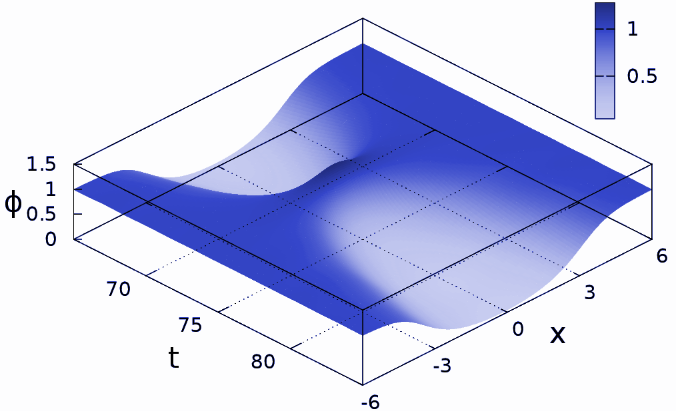}\label{fig:phi_101}}
  \subfigure[total energy density]{\includegraphics[width=0.49\textwidth]{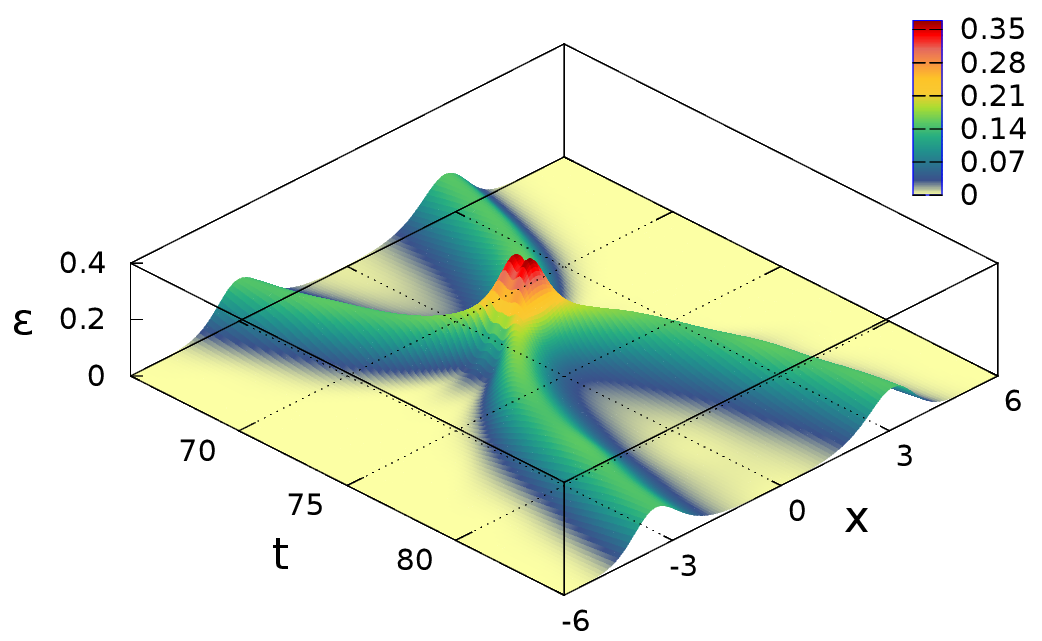}\label{fig:total_energy_101}}
  \\
  \subfigure[kinetic energy density]{\includegraphics[width=0.49\textwidth]{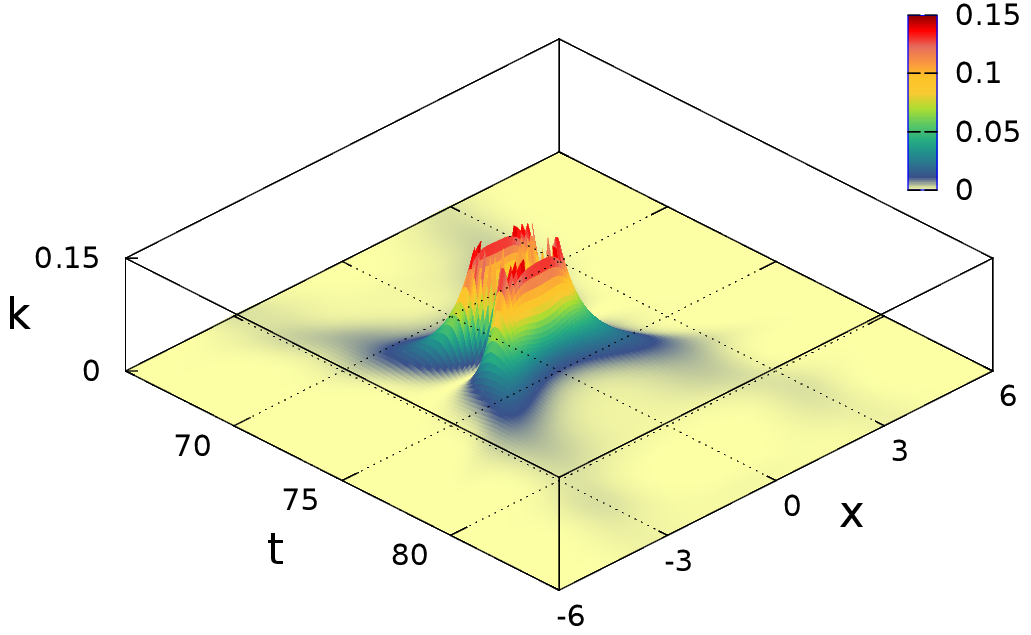}\label{fig:kinetic_energy_101}}
  \subfigure[potential energy density]{\includegraphics[width=0.49\textwidth]{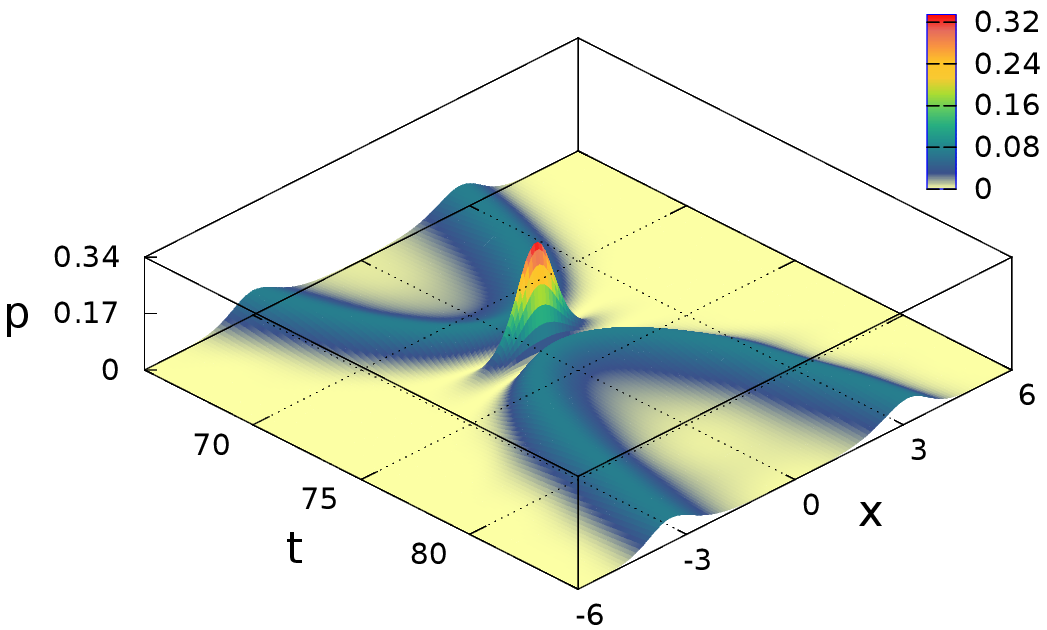}\label{fig:potential_energy_101}}
  \\
  \subfigure[gradient energy density]{\includegraphics[width=0.49\textwidth]{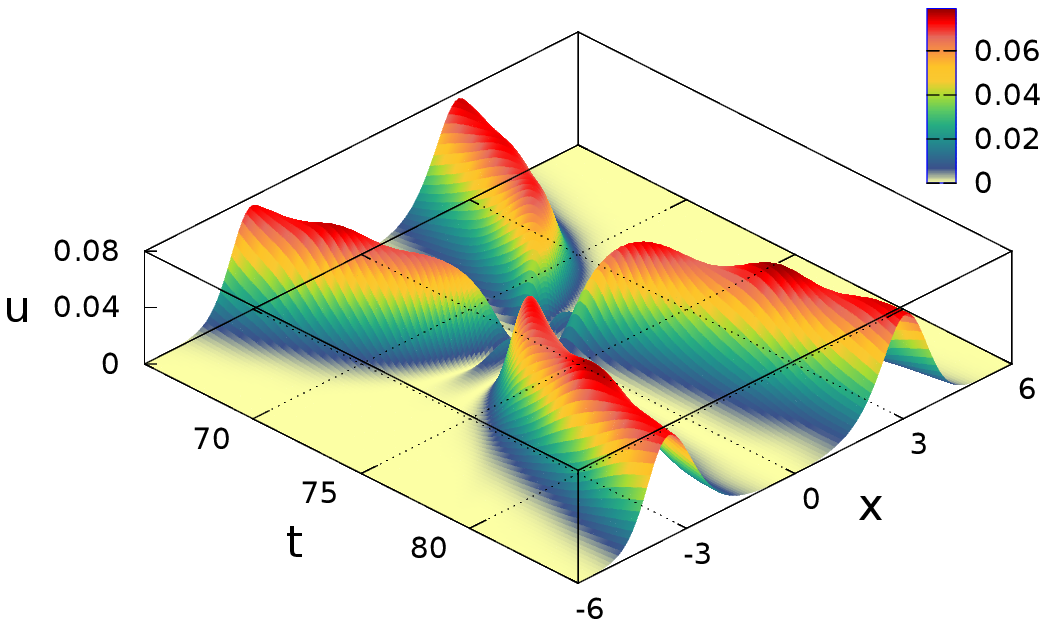}\label{fig:gradient_energy_101}}
  \subfigure[field gradient]{\includegraphics[width=0.49\textwidth]{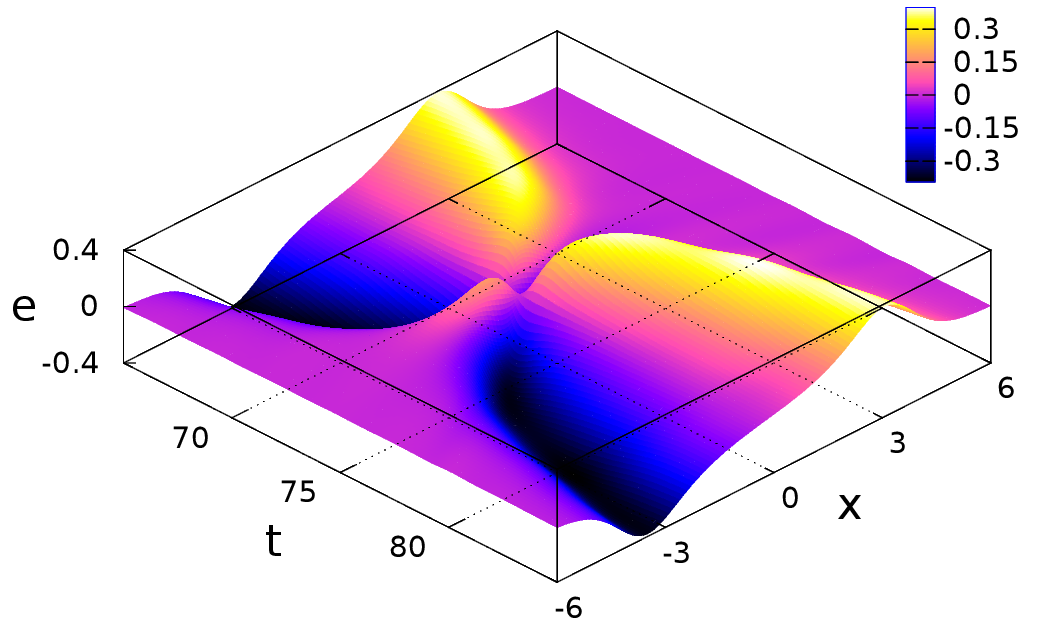}\label{fig:elastic_strain_101}}
  \caption{Space-time picture of collision of two kinks in the case of initial configuration $(1,0,1)$.}
  \label{fig:101}
\end{figure}

In figure \ref{fig:101} we give the results of our numerical simulation.
Figure \ref{fig:phi_101} shows the field profile before, during, and after the collision, illustrating their convergence, interaction during the collision, and escape. Moreover, figure \ref{fig:total_energy_101} shows that the kink and the antikink attract each other at small distances.

From the numerical analysis we obtain the following maximal values of the energy densities:
\begin{equation}
k_\mathrm{max}^{(2)} \approx 0.37,
\quad
u_\mathrm{max}^{(2)} \approx 0.07,
\quad
p_\mathrm{max}^{(2)} \approx 0.34,
\quad
\varepsilon_\mathrm{max}^{(2)} \approx 0.37.
\end{equation}
For the field gradient we have
\begin{equation}
e_\mathrm{min}^{(2)} \approx -0.4,
\quad
e_\mathrm{max}^{(2)} \approx 0.4.
\end{equation}
We see that the maximal values of the energy densities in the case of the configuration $(1,0,1)$ differ from those found in the case of the configuration $(0,1,0)$, while the extreme values of the field gradient are the same. The latter fact is explained by noticing that the maximal absolute values of the field gradient are observed in running non-interacting kinks. They can be calculated analytically from eq.~\eqref{eq:strain1}, which gives the value $e_\mathrm{max}^{(1)}\approx 0.3868$ for the single kink moving with the velocity $v=0.1$. 

We have also carried out a numerical simulation of the kink-antikink collision for the configuration $(-1,0,-1)$, and obtained the same results as for $(1,0,1)$.

\subsection{Collision of three kinks}\label{Sec:N3}

\subsubsection{The configuration (0,1,0,1)}

Our next step is to study collisions of three kinks at the same point. We start with the configuration of the type $(0,1,0,1)$:
\begin{equation}\label{Kink0101}
\phi_{(0,1,0,1)}(x,t) = \phi_{(0,1)}(x-x_1,t)+\phi_{(1,0)}(x-x_2,t)+\phi_{(0,1)}(x-x_3,t)-1.
\end{equation}
This configuration is composed of two kinks $(0,1)$ and one antikink $(1,0)$, which is placed between the kinks. We set the antikink to be static, $v_2=0$, while the kinks are moving towards it from the left and from the right with velocities $v_1=0.1$ and $v_3=-0.1$, respectively. The initial positions of the kinks are $x_1=-10$ and $x_3=10$. It turns out that, in order to ensure the simultaneous arrival of the two kinks at the location of the antikink, the latter has to be slightly shifted towards the left kink because of the asymmetry of the $\phi^6$ kinks. We found that the antikink must be placed at $x_2=-2.05954$ in order to obtain the maximal energy densities during the collision.

\begin{figure}[h!]
  \centering
  \subfigure[three kinks collision in the sector $(0,1,0,1)$]{\includegraphics[width=0.49\textwidth]{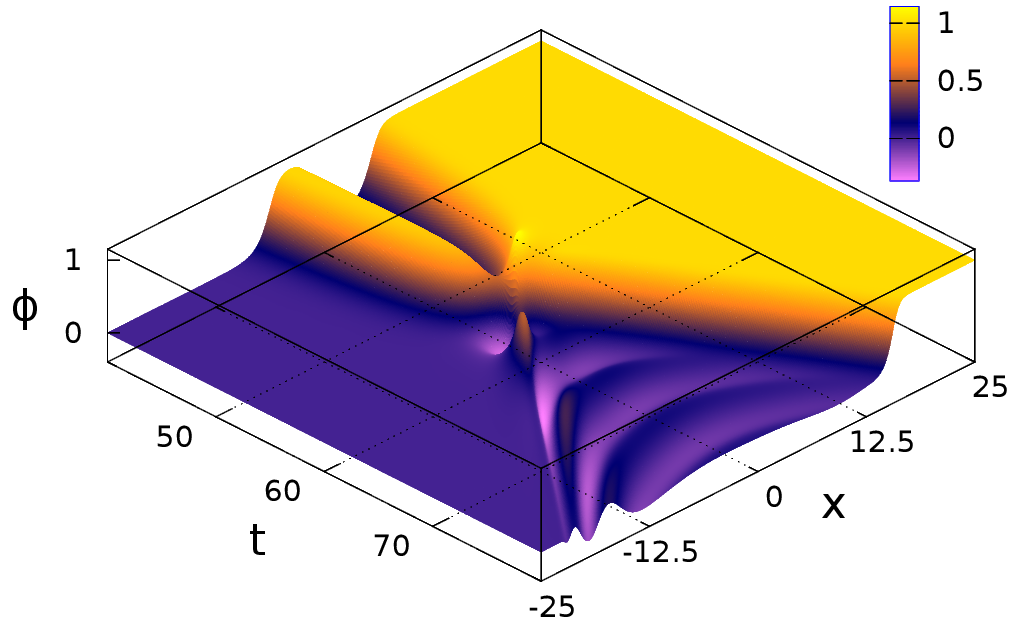}\label{fig:phi_0101}}
  \subfigure[total energy density]{\includegraphics[width=0.49\textwidth]{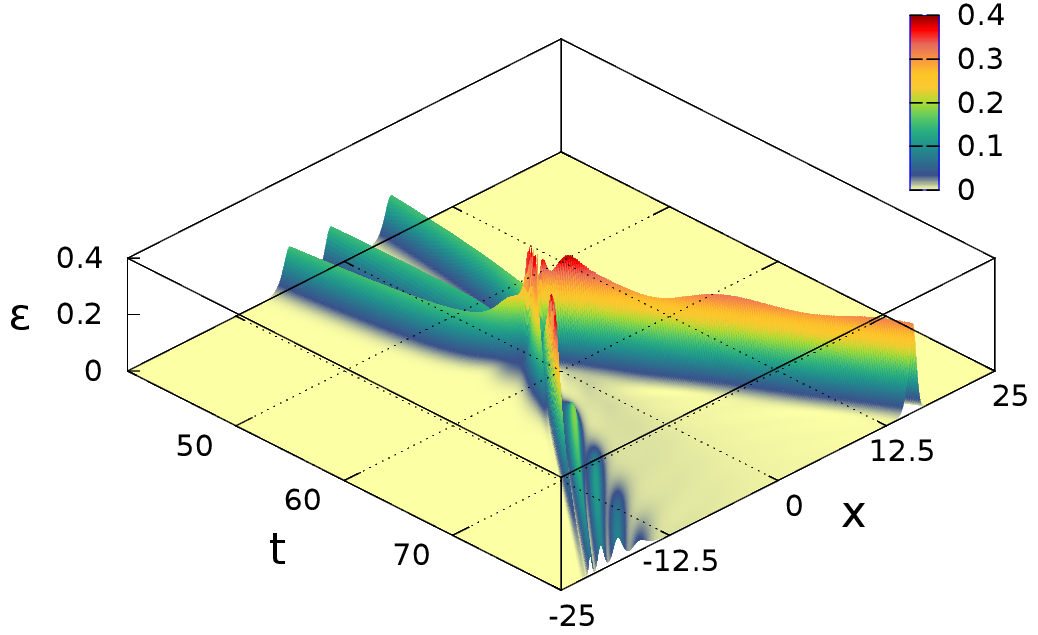}\label{fig:total_energy_0101}}
  \\
  \subfigure[kinetic energy density]{\includegraphics[width=0.49\textwidth]{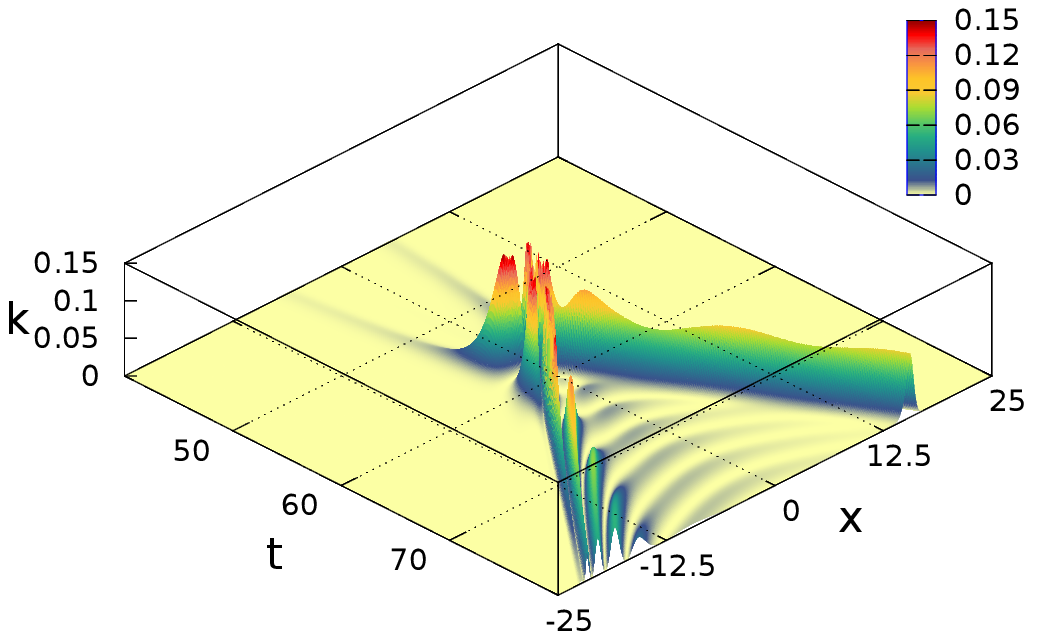}\label{fig:kinetic_energy_0101}}
  \subfigure[potential energy density]{\includegraphics[width=0.49\textwidth]{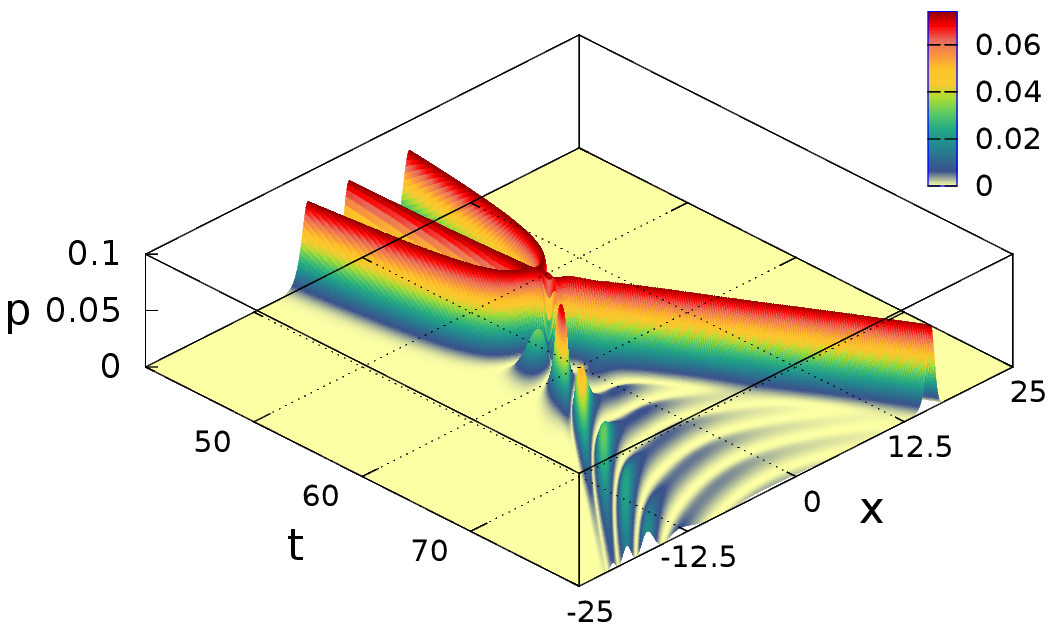}\label{fig:potential_energy_0101}}
  \\
  \subfigure[gradient energy density]{\includegraphics[width=0.49\textwidth]{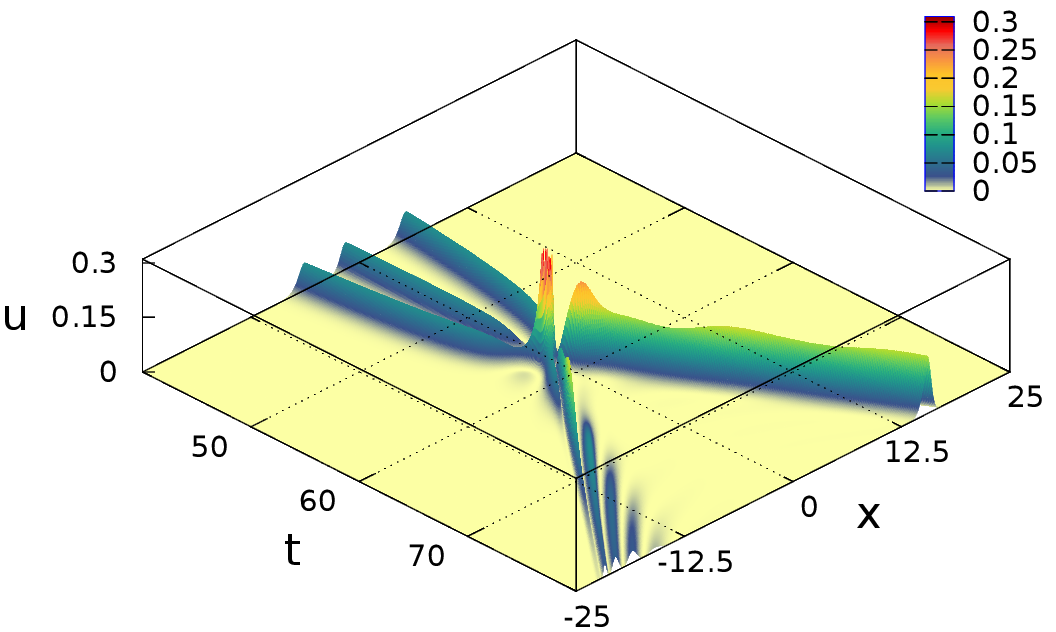}\label{fig:gradient_energy_0101}}
  \subfigure[field gradient]{\includegraphics[width=0.49\textwidth]{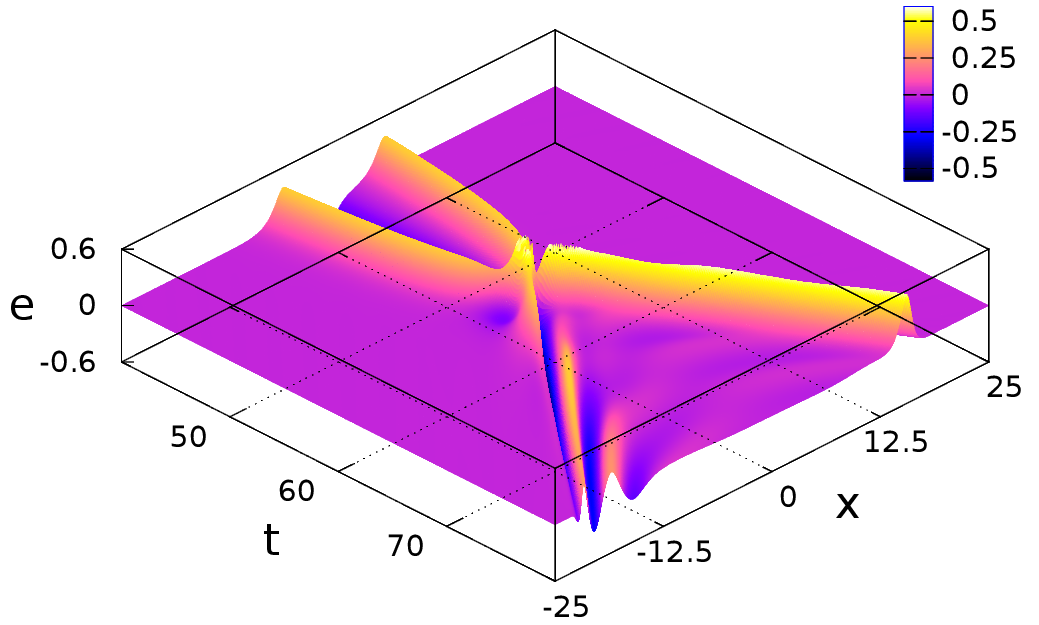}\label{fig:elastic_strain_0101}}
  \caption{Space-time picture of collision of three kinks in the case of initial configuration $(0,1,0,1)$.}
  \label{fig:0101}
\end{figure}

In figure \ref{fig:0101} we give the results of our numerical simulation of the kink-antikink-kink collision in the sector $(0,1,0,1)$.

The kink $(0,1)$ that started at $x=x_1$ and the antikink $(1,0)$, which are initially parts of the configuration of the type $(0,1,0)$, annihilate. They form an oscillating lump, which moves with a high velocity away from the collision point and quickly decays. At the same time, the other kink $(0,1)$, which originally started from the point $x_3$, survives and moves backwards after the collision. So we observe the ``reaction'' $K\bar{K}K\to bK$.

It is interesting to notice that the velocities of the lump and the kink after the collision substantially exceed the initial velocities of the colliding kinks. Apparently it is a consequence of redistribution of energy. When the kink and the antikink annihilate, a part of their energy is transferred to the kinetic energy of the surviving kink, increasing its speed.

We obtain the following extreme values of the energy densities and the field gradient:
\begin{equation}
k_\mathrm{max}^{(3)} \approx 0.27,
\quad
u_\mathrm{max}^{(3)} \approx 0.67,
\quad
p_\mathrm{max}^{(3)} \approx 0.075,
\quad
\varepsilon_\mathrm{max}^{(3)} \approx 0.75,
\end{equation}
and
\begin{equation}
e_\mathrm{min}^{(3)} \approx -0.58,
\quad
e_\mathrm{max}^{(3)} \approx 1.15.
\end{equation}

We have also carried out a numerical simulation of the antikink-kink-antikink collision for the configuration $(0,-1,0,-1)$ and obtained the same results as for $(0,1,0,1)$.

\subsubsection{The configuration (1,0,1,0)}

\begin{figure}[h!]
  \centering
  \subfigure[three kinks collision in the sector $(1,0,1,0)$]{\includegraphics[width=0.49\textwidth]{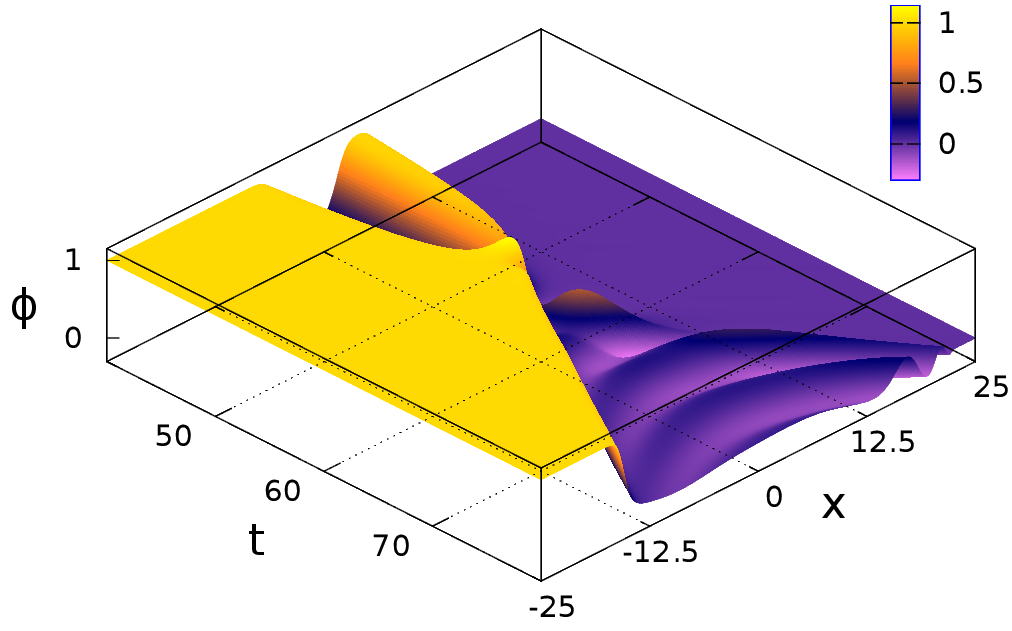}\label{fig:phi_1010}}
  \subfigure[total energy density]{\includegraphics[width=0.49\textwidth]{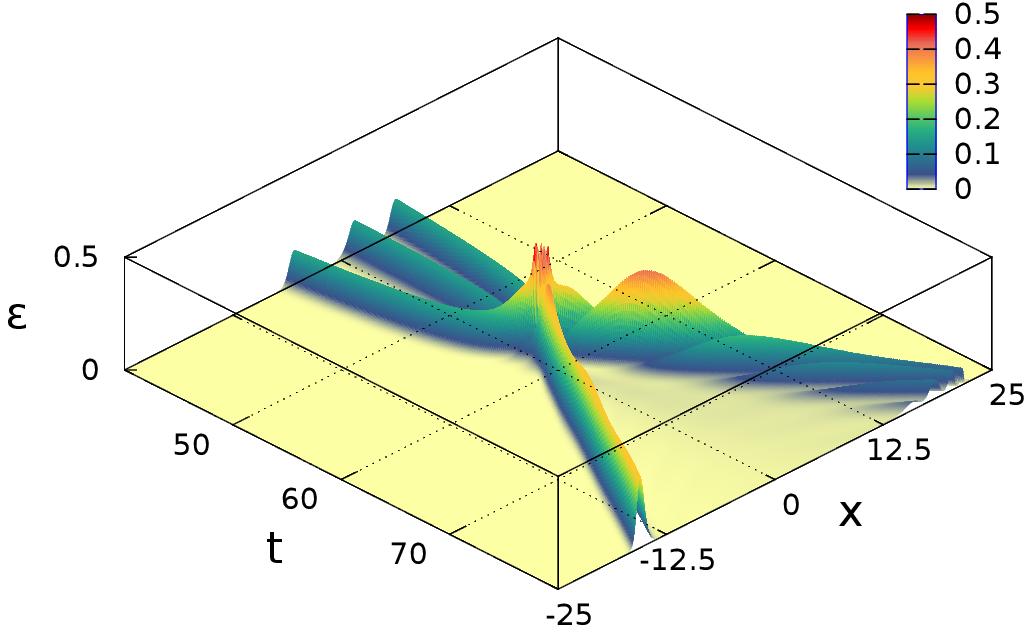}\label{fig:total_energy_1010}}
  \\
  \subfigure[kinetic energy density]{\includegraphics[width=0.49\textwidth]{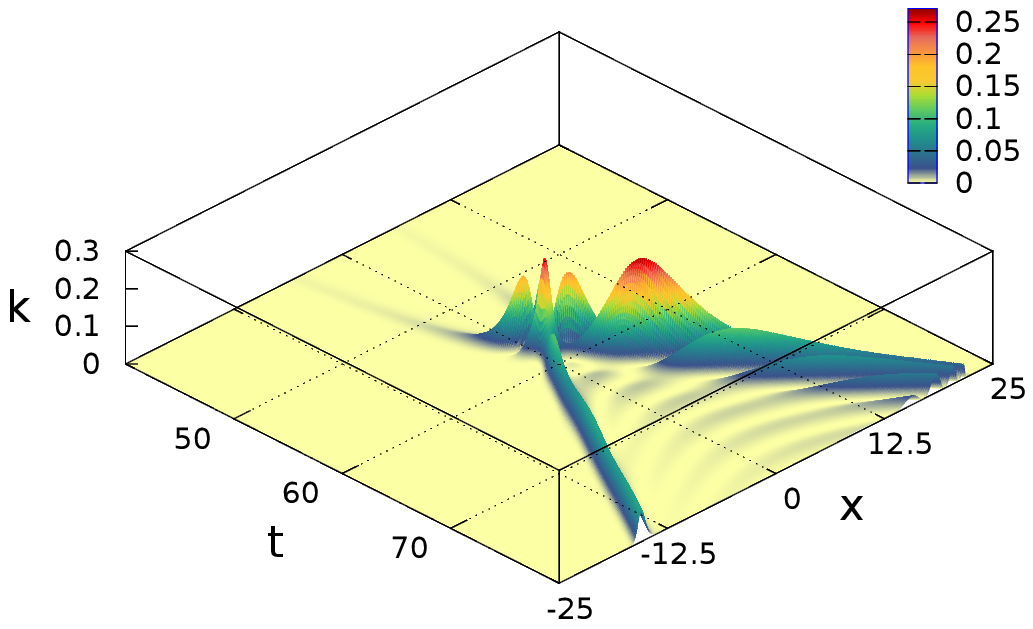}\label{fig:kinetic_energy_1010}}
  \subfigure[potential energy density]{\includegraphics[width=0.49\textwidth]{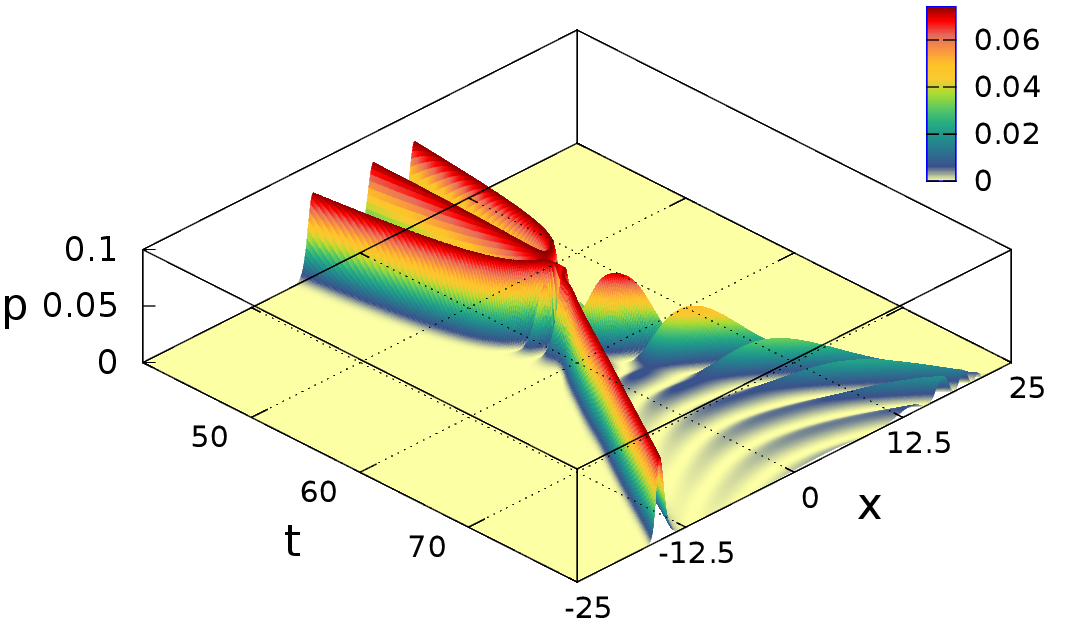}\label{fig:potential_energy_1010}}
  \\
  \subfigure[gradient energy density]{\includegraphics[width=0.49\textwidth]{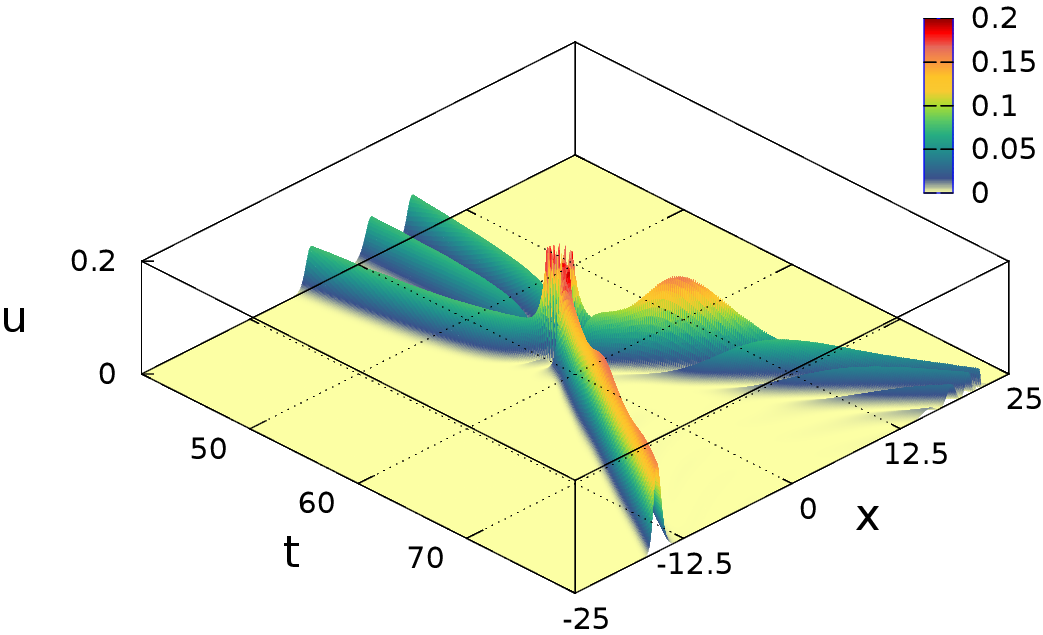}\label{fig:gradient_energy_1010}}
  \subfigure[field gradient]{\includegraphics[width=0.49\textwidth]{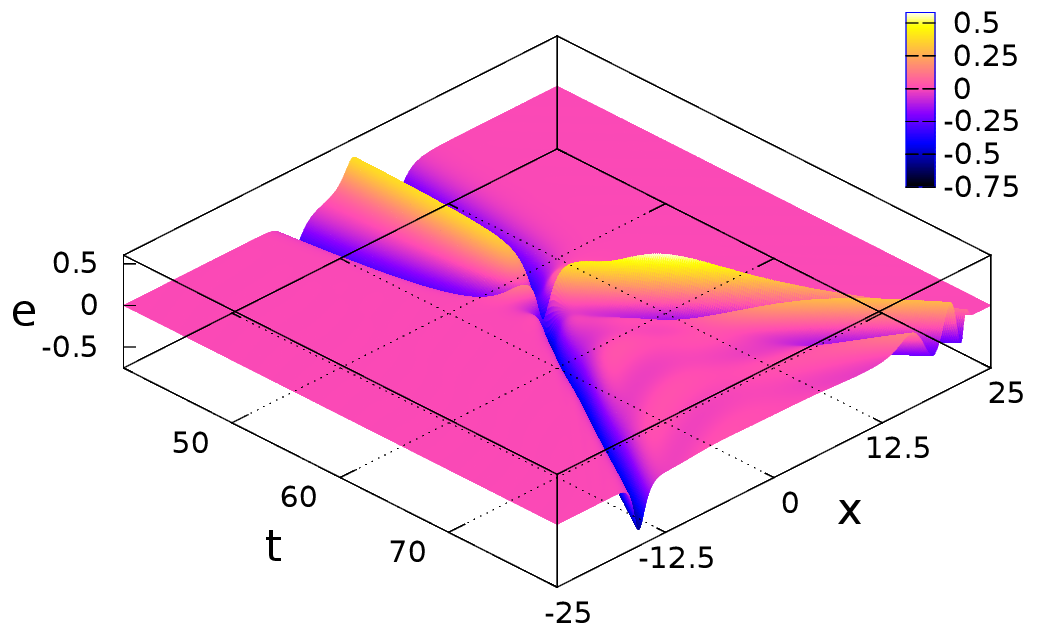}\label{fig:elastic_strain_1010}}
  \caption{Space-time picture of collision of three kinks in the case of initial configuration $(1,0,1,0)$.}
  \label{fig:1010}
\end{figure}

As in the previous subsection, we consider here a collision of three kinks, but with the initial configuration of the type $(1,0,1,0)$, which corresponds to two antikinks $(1,0)$ and one kink $(0,1)$ between them, i.e.\ $\bar{K}K\bar{K}$. For the numerical simulation we use the following initial condition:
\begin{equation}\label{Kink1010}
\phi_{(1,0,1,0)}(x,t) = \phi_{(1,0)}(x-x_1,t)+\phi_{(0,1)}(x-x_2,t)+\phi_{(1,0)}(x-x_3,t)-1.
\end{equation}
The antikinks are initially placed at $x_1=-10$ and $x_3=10$, and are moving towards each other with velocities $v_1=0.1$ and $v_3=-0.1$, while the static kink is initially located at $x_2=2.06000$, see figure \ref{fig:phi_1010}. As in the previous case, the position $x_2$ of the central kink ensures that both antikinks arrive at the location of the kink at the same time.

The results of the numerical simulation are shown in figure \ref{fig:1010}.
The collision pattern is quite similar to that of the configuration $(0,1,0,1)$. We observe the annihilation of the kink and one of the antikinks, which are initially parts of the configuration $(0,1,0)$. After annihilation they form an oscillating lump, which escapes from the collision point with a near-light speed. The other antikink $(1,0)$, which survives in the collision, also escapes with a near-light speed --- its final velocity substantially exceeds the initial velocity. The observed antikink-kink-antikink collision is the ``reaction'' $\bar{K}K\bar{K}\to\bar{K}b$.

From the numerical analysis we extract the following extreme values:
\begin{equation}
k_\mathrm{max}^{(3)} \approx 0.27,
\quad
u_\mathrm{max}^{(3)} \approx 0.67,
\quad
p_\mathrm{max}^{(3)} \approx 0.075,
\quad
\varepsilon_\mathrm{max}^{(3)} \approx 0.75,
\end{equation}
and
\begin{equation}
e_\mathrm{min}^{(3)} \approx -1.15,
\quad
e_\mathrm{max}^{(3)} \approx 0.57.
\end{equation}
From this results we see that the maximal values of the energy densities are the same as for the configuration $(0,1,0,1)$, while the extreme values of the field gradient are different.

We have also carried out numerical simulation of the kink-antikink-kink collision for the configuration $(-1,0,-1,0)$ and obtained the same results as for $(1,0,1,0)$.

\subsection{Collision of four kinks}
\label{Sec:N4}

\subsubsection{The configuration (0,1,0,1,0)}

\begin{figure}[h!]
  \centering
  \subfigure[four kinks collision in the sector $(0,1,0,1,0)$]{\includegraphics[width=0.49\textwidth]{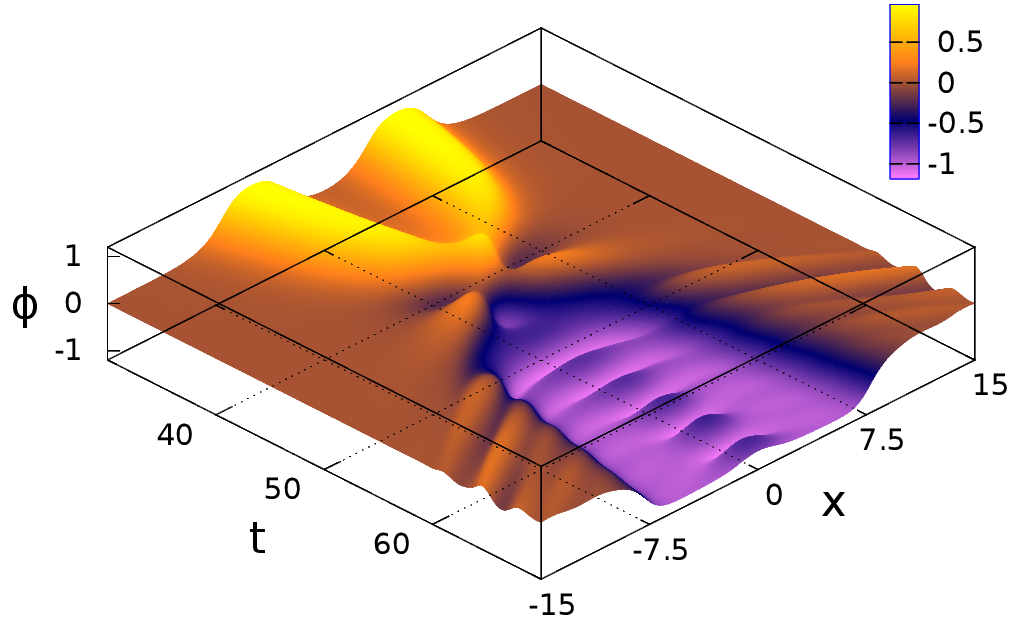}\label{fig:phi_01010}}
  \subfigure[total energy density]{\includegraphics[width=0.49\textwidth]{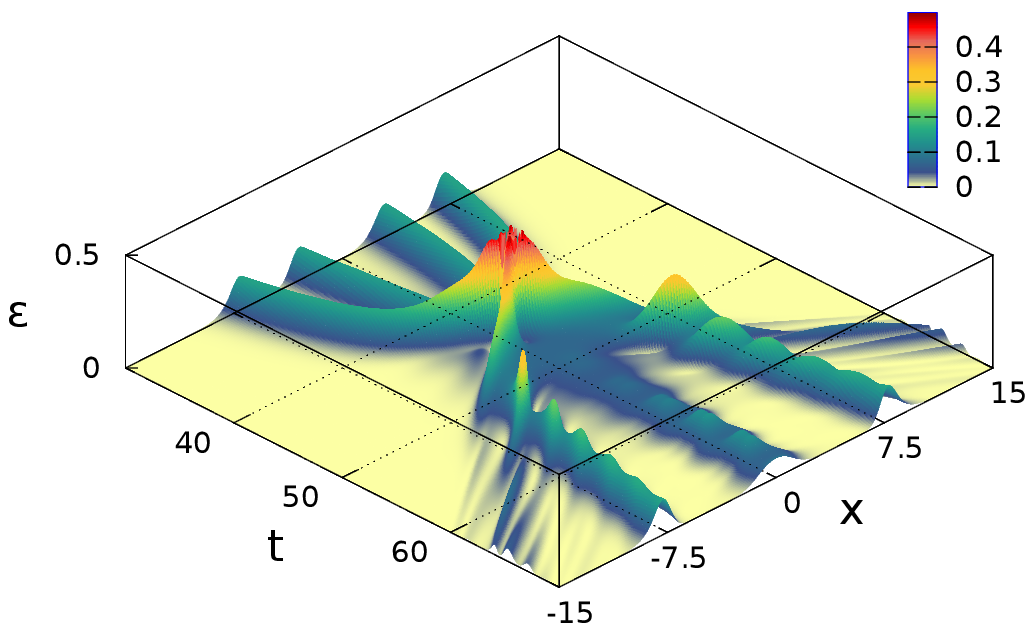}\label{fig:total_energy_01010}}
  \\
  \subfigure[kinetic energy density]{\includegraphics[width=0.49\textwidth]{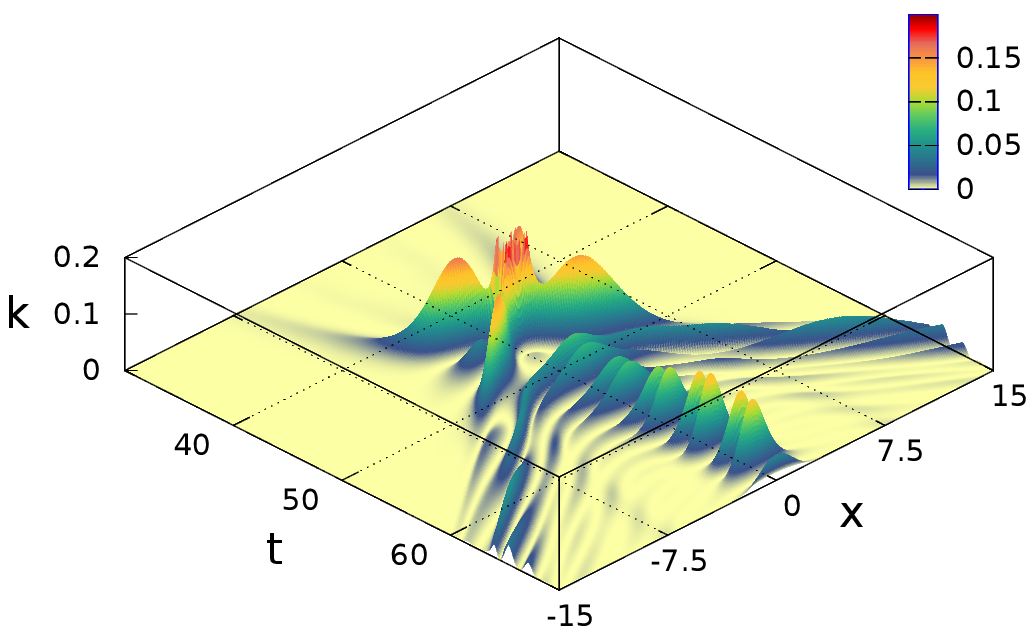}\label{fig:kinetic_energy_01010}}
  \subfigure[potential energy density]{\includegraphics[width=0.49\textwidth]{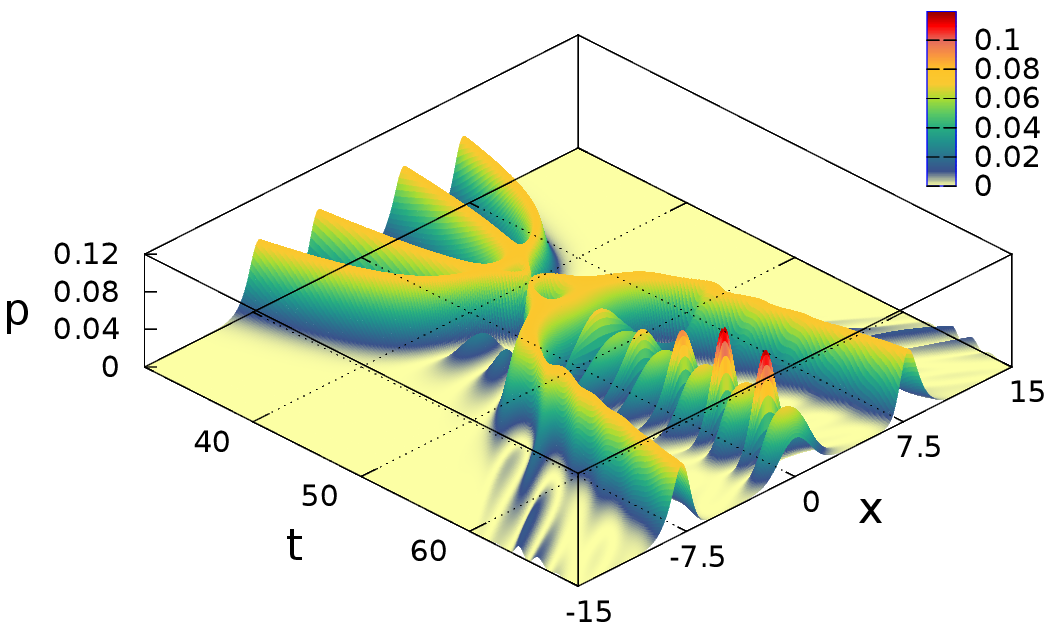}\label{fig:potential_energy_01010}}
  \\
  \subfigure[gradient energy density]{\includegraphics[width=0.49\textwidth]{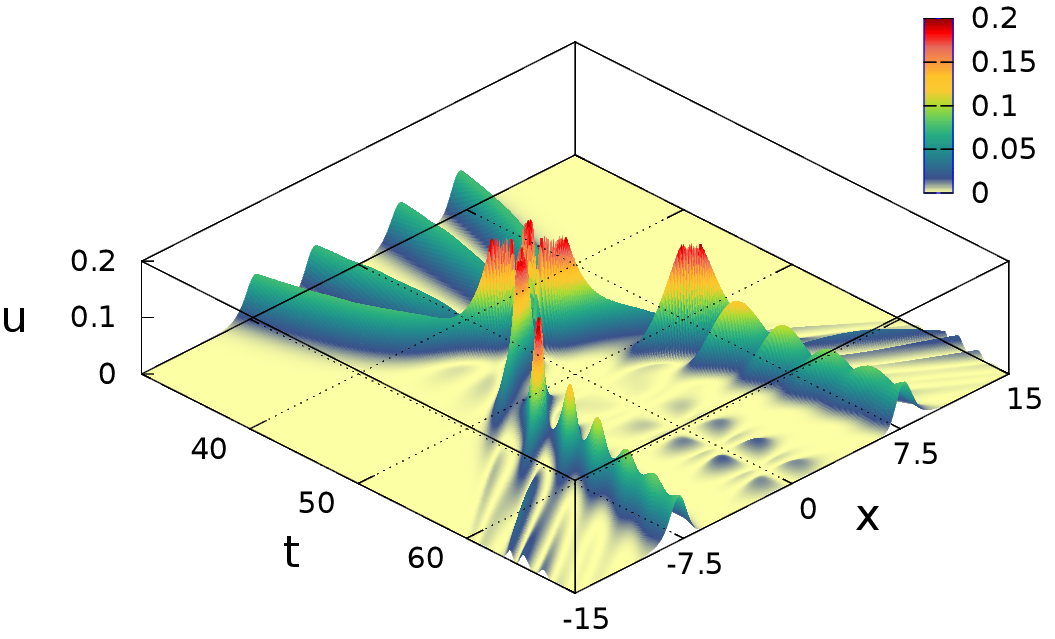}\label{fig:gradient_energy_01010}}
  \subfigure[field gradient]{\includegraphics[width=0.49\textwidth]{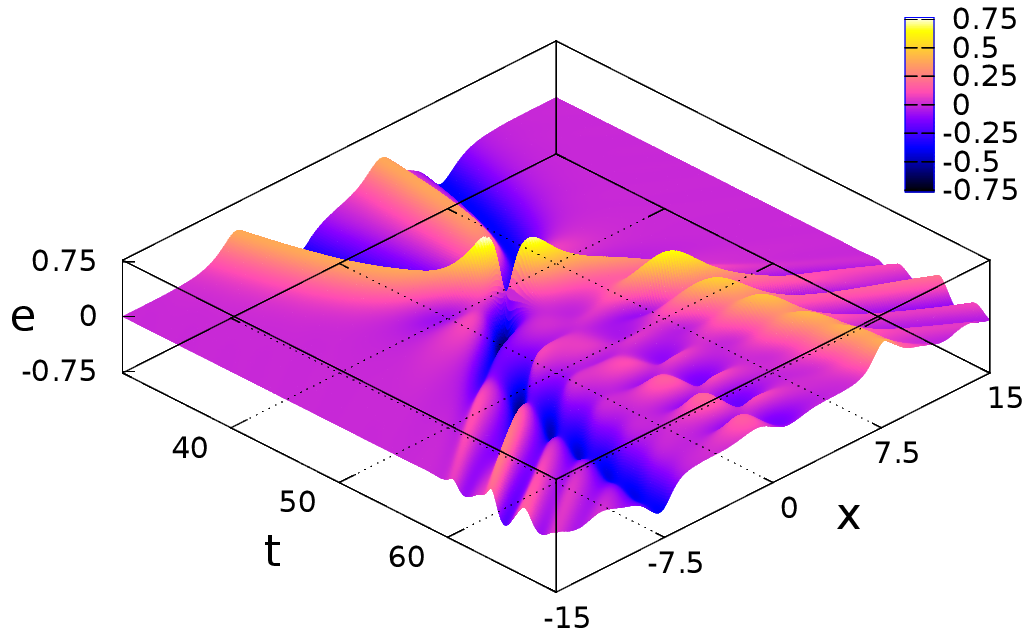}\label{fig:elastic_strain_01010}}
  \caption{Space-time picture of collision of four kinks in the case of initial configuration $(0,1,0,1,0)$.}
  \label{fig:01010}
\end{figure}

We move on to consider collisions of four kinks and antikinks. We start with the following initial configuration:
\begin{equation}\label{Kink01010}
\phi_{(0,1,0,1,0)}(x,t) = \phi_{(0,1)}(x-x_1,t)+\phi_{(1,0)}(x-x_2,t)+\phi_{(0,1)}(x-x_3,t)+\phi_{(1,0)}(x-x_4,t)-2,
\end{equation}
where $x_1=-x_4=-10.17604$, $v_1=-v_4=0.1$, $x_2=-x_3=-5.0$, $v_2=-v_3=0.05$. At these initial conditions the collision of all four solitons occurs at the same point, see figure \ref{fig:01010}.
After the collision we observe the formation of a bion at the origin (the collision point), and an antikink $(0,-1)$ with a kink $(-1,0)$, which are moving with constant velocities in the opposite directions away from the collision point. So we have a process of the type $K\bar{K}K\bar{K}\to \bar{K}bK$. Of course, some energy in the form of waves of small amplitude is emitted during the collision. The observed reaction can be interpreted as follows. One kink-antikink pair forms a bound state (bion), while the other pair scatters, and in the final state we observe a configuration of the type $(0,-1,0)$ with the bion at its center, see figure \ref{fig:phi_01010}.

We found the following extreme values of the energy densities and the field gradient for this collision of four kinks:
\begin{equation}
k_\mathrm{max}^{(4)} \approx 0.95,
\quad
u_\mathrm{max}^{(4)} \approx 0.3,
\quad
p_\mathrm{max}^{(4)} \approx 0.075,
\quad
\varepsilon_\mathrm{max}^{(4)} \approx 0.95,
\end{equation}
and
\begin{equation}
e_\mathrm{min}^{(4)} \approx -0.77,
\quad
e_\mathrm{max}^{(4)} \approx 0.77.
\end{equation}

\subsubsection{The configuration (1,0,1,0,1)}

Finally, we consider the initial configuration of the type $(1,0,1,0,1)$, i.e., the $\bar{K}K\bar{K}K$ system:
\begin{equation}\label{Kink10101}
\phi_{(1,0,1,0,1)}(x,t) = \phi_{(1,0)}(x-x_1,t)+\phi_{(0,1)}(x-x_2,t)+\phi_{(1,0)}(x-x_3,t)+\phi_{(0,1)}(x-x_4,t)-1,
\end{equation}
where $x_1=-x_4=-17.1375468$, $v_1=-v_4=0.1$, $x_2=-x_3=-5.0$, $v_2=-v_3=0.05$. As in the previous cases, we use specially chosen initial positions and initial velocities of the kinks and antikinks in order to force them all to collide at the same point.

\begin{figure}[h!]
  \centering
  \subfigure[four kinks collision in the sector $(1,0,1,0,1)$]{\includegraphics[width=0.49\textwidth]{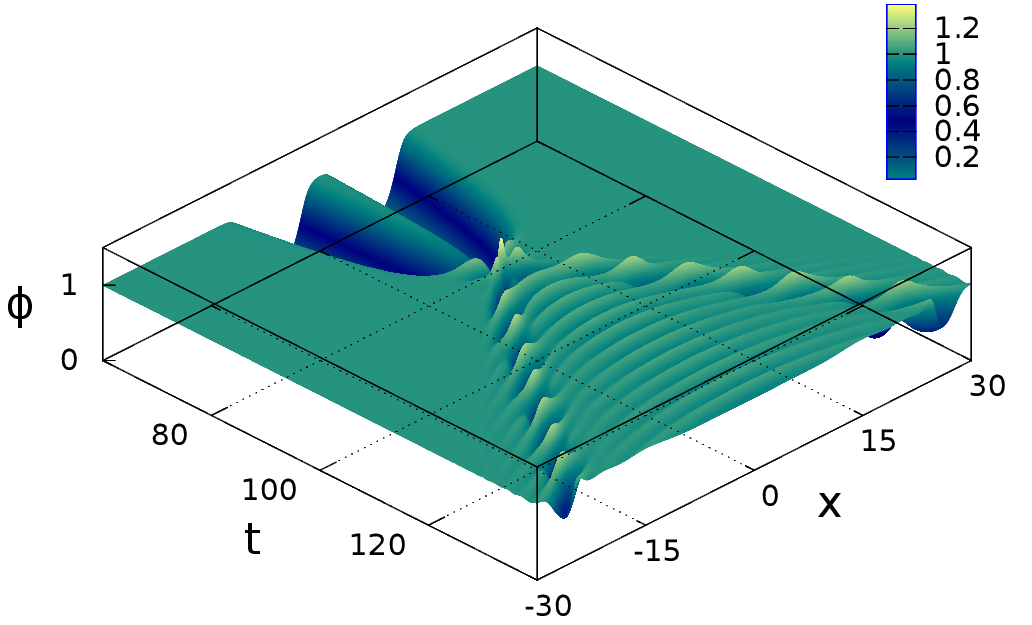}\label{fig:phi_10101}}
  \subfigure[total energy density]{\includegraphics[width=0.49\textwidth]{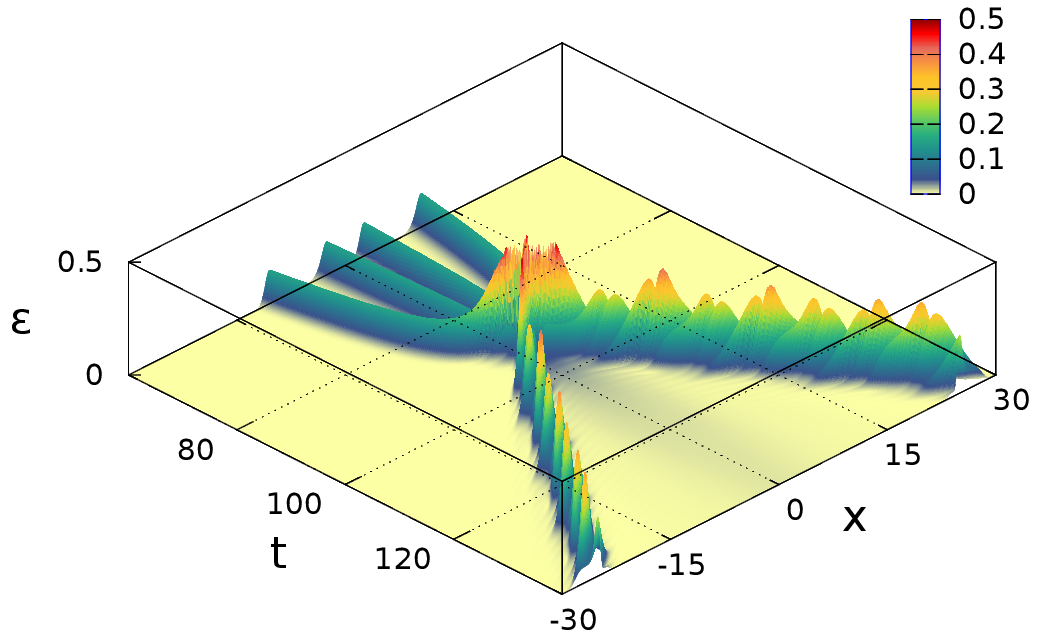}\label{fig:total_energy_10101}}
  \\
  \subfigure[kinetic energy density]{\includegraphics[width=0.49\textwidth]{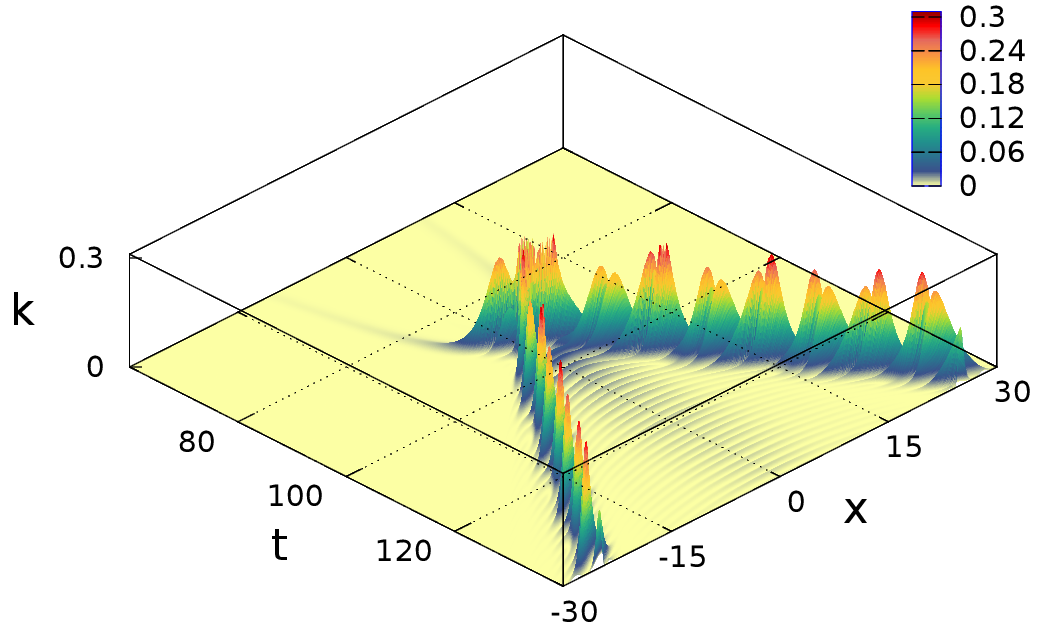}\label{fig:kinetic_energy_10101}}
  \subfigure[potential energy density]{\includegraphics[width=0.49\textwidth]{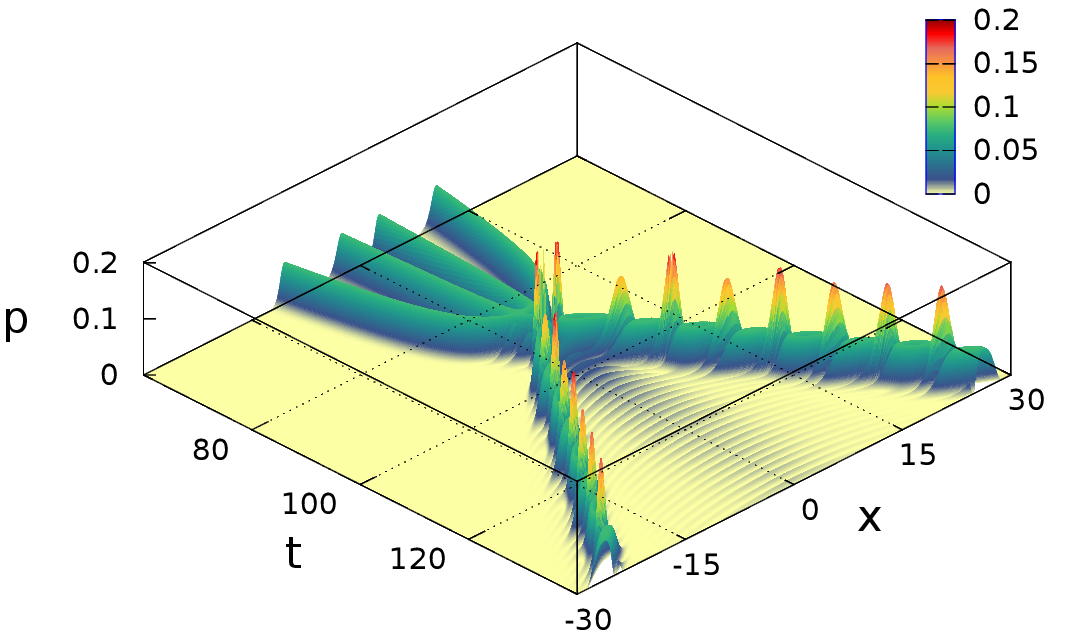}\label{fig:potential_energy_10101}}
  \\
  \subfigure[gradient energy density]{\includegraphics[width=0.49\textwidth]{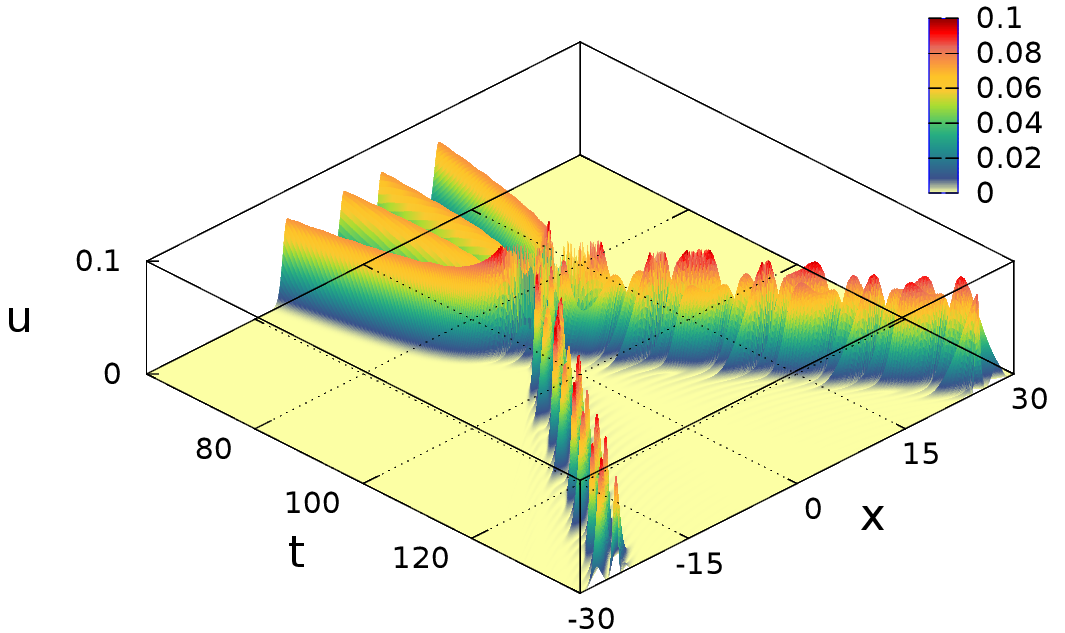}\label{fig:gradient_energy_10101}}
  \subfigure[field gradient]{\includegraphics[width=0.49\textwidth]{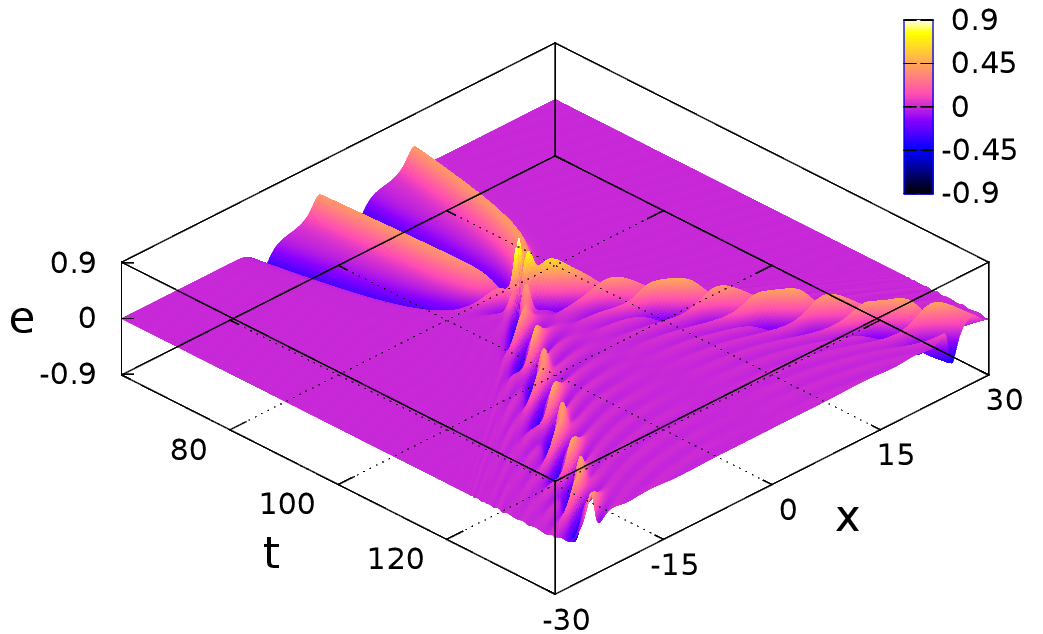}\label{fig:elastic_strain_10101}}
  \caption{Space-time picture of collision of four kinks in the case of initial configuration $(1,0,1,0,1)$.}
  \label{fig:10101}
\end{figure}

The results of the numerical simulation are shown in figure \ref{fig:10101}.

In particular, from figures \ref{fig:phi_10101} and \ref{fig:total_energy_10101} it is clear that two $K\bar{K}$ bound states (bions) are formed after the collision, so we have a process of the type $\bar{K}K\bar{K}K\to bb$. The bions escape from the collision point with velocities that substantially exceed the initial velocities of the colliding kinks. Notice that the situation is different from that we observed in the case of the initial configuration $(0,1,0,1,0)$, described in the previous subsection.

For the extreme values of the energy densities and the field gradient we find:
\begin{equation}
k_\mathrm{max}^{(4)} \approx 1.47,
\quad
u_\mathrm{max}^{(4)} \approx 0.42,
\quad
p_\mathrm{max}^{(4)} \approx 0.78,
\quad
\varepsilon_\mathrm{max}^{(4)} \approx 1.47,
\end{equation}
and
\begin{equation}
e_\mathrm{min}^{(4)} \approx -0.91,
\quad
e_\mathrm{max}^{(4)} \approx 0.91.
\end{equation}

\section{Conclusion}\label{sec:Conclusion}

We have studied the process of collision of several $\phi^6$ kinks and antikinks at the same point. We used the initial configurations of the following types: $K\bar{K}$, $\bar{K}K$, $K\bar{K}K$, $\bar{K}K\bar{K}$, $K\bar{K}K\bar{K}$, and $\bar{K}K\bar{K}K$. In all these cases the initial positions and initial velocities were fitted so as to achieve the simultaneous collision of all solitons at the same point. For each initial configuration we restricted ourselves to only one set of the initial data. The results are collected in tables \ref{tab:Table1} and \ref{tab:Table2}.

\begin{table}
\caption{Initial and final velocities of quasiparticles.}
\label{tab:Table1}
\begin{tabular}{|c|c|c|c|c|c|}
\hline
$N$ & type & initial state  & initial velocities& final state & final velocities\\
\hline
2\vphantom{$\displaystyle\frac{1}{2}$} & $(0,1,0)$ & $K\bar{K}$ & $v_K=0.1,v_{\bar{K}}=-0.1$ & $b$ & $v_b=0$\\
\hline
2\vphantom{$\displaystyle\frac{1}{2}$}  & $(1,0,1)$ & $\bar{K}K$ & $v_{\bar{K}}=0.1$, $v_K=-0.1$ & $\bar{K}K$ & $v_K=-v_{\bar{K}}\approx 0.1$\\
\hline
3\vphantom{$\displaystyle\frac{1}{2}$} & $(0,1,0,1)$ & $K_1\bar{K}K_2$ & $v_{K_1}=-v_{K_2}=0.1$, & $bK$ &$v_b\approx -0.9$,\\
 & & & $v_{\bar{K}}=0$ & & $v_K\approx 0.7$\\
\hline
3\vphantom{$\displaystyle\frac{1}{2}$} & $(1,0,1,0)$ & $\bar{K}_1K\bar{K}_2$ &$v_{\bar{K}_1}=-v_{\bar{K}_2}=0.1$, & $\bar{K}b$ & $v_{\bar{K}}\approx -0.7$,\\
 & & & $v_K=0$ & & $v_b\approx 0.9$\\
\hline
4\vphantom{$\displaystyle\frac{1}{2}$} & $(0,1,0,1,0)$ & $K_1\bar{K}_1K_2\bar{K}_2$ & $v_{K_1}=-v_{\bar{K}_2}=0.1$, & $\bar{K}bK$ & $v_K\approx -v_{\bar{K}}\approx 0.13$,\\
 & & & $v_{\bar{K}_1}=-v_{K_2}=0.05$ & & $v_b=0$\\
\hline
4\vphantom{$\displaystyle\frac{1}{2}$} & $(1,0,1,0,1)$ & $\bar{K}_1K_1\bar{K}_2K_2$ & $v_{\bar{K}_1}=-v_{K_2}=0.1$, & $b_1\:b_2$ & $v_{b_2} \approx -v_{b_1} \approx 0.5$\\
 & & & $v_{K_1}=-v_{\bar{K}_2}=0.05$ & & \\
\hline
\end{tabular}
\end{table}

In table \ref{tab:Table1} we give the initial velocities of the kinks and antikinks together with the final velocities of the quasiparticles for all collisions discussed in this work. Depending on the number of the kinks and on their location order in the initial configuration, we observed reflection, passing through each other, and capture. Apparently, a particular final state configuration is a consequence of the rather complicated picture of the pairwise kink-antikink interactions: the two solitons can form a bound state or reflect off each other, depending on the value of their initial velocities. The critical velocities that separate these two regimes also depend on the type of the initial configuration, $K\bar{K}$ or $\bar{K}K$, because the kinks of the $\phi^6$ model are not symmetric: they have different spatial asymptotics depending on the vacuum to which the field $\phi$ tends, see eqs.~\eqref{eq:asymptotics1}, \eqref{eq:asymptotics2}.

In the case of the kink-antikink collisions ($N=2$) we reproduced the well-known scenarios. In the $K\bar{K}$ collision [the initial configuration of the type $(0,1,0)$] at $v_\mathrm{in}=0.1$ we observed the capture of the kink and the antikink and the formation of a bion --- a kink-antikink bound state. This happens because $v_\mathrm{in}<v_\mathrm{cr}\approx 0.289$. At the same time, in the $\bar{K}K$ collision [the initial configuration of the type $(1,0,1)$] at $v_\mathrm{in}=0.1$ we could see the two solitons to escape after the collision. Such behavior is a consequence of the fact that now $v_\mathrm{in}>v_\mathrm{cr}\approx 0.045$.

Next, we studied collisions of three solitons, namely two kinks and one antikink [the initial configuration of the type$(0,1,0,1)$, or $K\bar{K}K$] and of two antikinks and one kink [the initial configuration of the type $(1,0,1,0)$, or $\bar{K}K\bar{K}$]. The observed final states in these cases seem to be due to the kink-antikink capture at low energies, $K\bar{K}\to b$, see table \ref{tab:Table1}.

In the case $N=4$ the situation is more complicated. On the one hand, in the process $K\bar{K}K\bar{K}\to \bar{K}bK$ we observed two features: a) the formation of a bion, and b) a kink and an antikink passing through each other and escaping to infinities. On the other hand, in the process $\bar{K}K\bar{K}K\to b\:b$ we have two bions in the final state, moving with high velocities in the opposite directions from the collision point. Thus in the latter case we observe annihilation of all four solitons.

\begin{table}
\caption{Extreme values of the energy densities and the field gradient observed in the studied kink collisions.}
\label{tab:Table2}
\begin{center}
\begin{tabular}{|c|c|c|c|c|c|c|c|}
\hline
$N$ & initial state & $\varepsilon_\mathrm{max}^{(N)}$ & $k_\mathrm{max}^{(N)}$ & $p_\mathrm{max}^{(N)}$ & $u_\mathrm{max}^{(N)}$ & $e_\mathrm{min}^{(N)}$ & $e_\mathrm{max}^{(N)}$ \\
\hline
1 & $K$ & 0.15 & 0.0007 & 0.074 & 0.075 & 0.0 & 0.4\\
2 & $K\bar{K}$ & 0.25& 0.25 & 0.075 & 0.075 & -0.4 & 0.4\\
2 & $\bar{K}K$ & 0.37 & 0.37 & 0.34 & 0.07 & -0.4 & 0.4\\
3 & $K\bar{K}K$ & 0.75 & 0.27 & 0.075 & 0.67 & -0.58 & 1.15\\
3 & $\bar{K}K\bar{K}$ & 0.75 & 0.27 & 0.075 & 0.67 & -1.15 & 0.57\\
4 & $K\bar{K}K\bar{K}$ & 0.95 & 0.95 & $0.12^*$ & 0.3 & -0.77 & 0.77\\
4 & $\bar{K}K\bar{K}K$ & 1.47 & 1.47 & 0.78 & 0.42 & -0.91 & 0.91\\
\hline
\end{tabular}
\end{center}
\end{table}

The extreme values of the energy densities and the field gradient are presented in table \ref{tab:Table2}. Recall that $\varepsilon_\mathrm{max}$, $k_\mathrm{max}$, $p_\mathrm{max}$, and $u_\mathrm{max}$ are the maximal densities of the total, kinetic, potential, and gradient energy, respectively, while $e_\mathrm{min}$ and $e_\mathrm{max}$ are the minimal and maximal values of the field gradient. The values in the first line of table \ref{tab:Table2} were calculated for a single kink with the help of analytic expressions \eqref{eq:energy1} and \eqref{eq:strain1} for the kink velocity $v=0.1$ (the same velocity as in the simulations of soliton collisions). We see that the maximal energy density for a single kink is 0.15 and that it is nearly equally shared between the potential and the gradient energy, while the maximal kinetic energy density is rather small at this velocity. A single kink produces the maximal tensile strain of 0.4, while the antikink yields the maximal compressive (negative) strain of the same magnitude. In the $K\bar{K}$ collisions, the maximal energy density is 0.25 and it is in the form of the kinetic energy density. The maximal (minimal) field gradient is the same as for a single kink (antikink). Due to the asymmetry of the $\phi^6$ kinks, the $\bar{K}K$ collisions produce somewhat higher maximal total energy density of 0.37, also in the form of the kinetic energy density. The extreme values of the field gradient are the same as for a single kink (antikink). In three-kink collisions the maximal energy density rises up to 0.75, which is five times larger than that of a single kink. The maximal and minimal values of the field gradient are 1.15 and $-1.15$, respectively, which is nearly three times larger than in a single kink. In the four-kink collision $K\bar{K}K\bar{K}$ the maximal energy density is 6.3 times larger than in a single kink. Strikingly, in the case of the $\bar{K}K\bar{K}K$ collision the maximal energy density is almost 10 times larger than in a single kink, and it is in the form of the kinetic energy. The extreme values of the field gradient in the four-kink collisions are roughly two times greater than in the case of a single kink. We thus conclude that, in multi-kink collisions, very high energy density spots can be observed. Notice that the maximum value of the potential energy density $p_\mathrm{max}=0.12$ in the process $K\bar{K}K\bar{K}\to \bar{K}bK$ we observe in the bion oscillations after the kinks' collision, see figure \ref{fig:potential_energy_01010}.

In conclusion, we emphasize that this work opens wide prospects for future research. In particular, it would be interesting to study multi-kink collisions within the $\phi^8$ model \cite{lohe,khare,GaLeLi,GaLeLiconf}. Depending on the parameters of the model, the $\phi^8$ kinks can have vibrational modes. These modes, in turn, can affect the energy redistribution in the multi-kink collisions. Besides that, the kinks of the $\phi^8$ model, corresponding to particular choices of the parameters, can have power-law asymptotics, which leads to a long-range interaction between kink and antikink. Therefore the multi-kink scattering can have new interesting features.

We would also like to notice that the multi-kink collisions can produce quasiparticles, which have very high speed. These quasiparticles of the ``second generation'' can, in turn, be forced to collide at the same point. Study of such processes can be a subject of future research.

\section*{Acknowledgments}

This research was supported by the MEPhI Academic Excellence Project (contract No.\ 02.a03.21.0005, 27.08.2013). S.V.D.\ thanks the Russian Science Foundation for their financial support under the grant N 16-12-10175. D.S.\ is grateful for the partial financial support provided by the Russian Science Foundation under the grant N 14-13-00982.


\begin{thebibliography}{99}

\bibitem{vilenkin01}
A.~Vilenkin and E.P.S.~Shellard, {\it Cosmic Strings and Other Topological Defects}, Cambridge University Press, Cambridge U.K. (2000).

\bibitem{manton01}
N.~Manton and P.~Sutcliffe, {\it Topological Solitons}, Cambridge University Press, Cambridge U.K. (2004).

\bibitem{aek01}
T.I.~Belova and A.E.~Kudryavtsev, {\it Solitons and their interactions in classical field theory}, \href{https://doi.org/10.1070/PU1997v040n04ABEH000227}{{\it Phys.~Usp.} {\bf 40} (1997) 359} [\href{https://doi.org/10.3367/UFNr.0167.199704b.0377}{{\it Usp.~Fiz.~Nauk} {\bf 167} (1997) 377}].

\bibitem{nitta1}
M.~Nitta, {\it Josephson vortices and the Atiyah-Manton construction}, \href{https://doi.org/10.1103/PhysRevD.86.125004}{{\it Phys.~Rev.} {\bf D 86} (2012) 125004} [\href{https://arxiv.org/abs/1207.6958}{\tt arXiv:1207.6958}].

\bibitem{nitta2}
M.~Nitta, {\it Correspondence between Skyrmions in 2+1 and 3+1 dimensions}, \href{https://doi.org/10.1103/PhysRevD.87.025013}{{\it Phys.~Rev.} {\bf D 87} (2013) 025013} [\href{https://arxiv.org/abs/1210.2233}{\tt arXiv:1210.2233}].

\bibitem{kur01}
E.~Kurianovych and M.~Shifman, {\it Non-Abelian moduli on domain walls}, \href{https://doi.org/10.1142/S0217751X14501930}{{\it Int.~J.~Mod.~Phys.} {\bf A 29} (2014) 1450193} [\href{https://arxiv.org/abs/1407.7144}{\tt arXiv:1407.7144}].

\bibitem{blyankinshtein}
N.~Blyankinshtein, {\it Q-lumps on a domain wall with a spin-orbit interaction}, \href{https://doi.org/10.1103/PhysRevD.93.065030}{{\it Phys.~Rev.} {\bf D 93} (2016) 065030} [\href{https://arxiv.org/abs/1510.07935}{\tt arXiv:1510.07935}].

\bibitem{kur02}
V.~Bychkov, M.~Kreshchuk and E.~Kurianovych, {\it More about structures localized on domain walls: strings, skyrmions, analytic solutions for orientational moduli, symmetry analysis}, \href{https://arxiv.org/abs/1603.06310}{\tt arXiv:1603.06310}.

\bibitem{nitta3}
M.~Nitta, {\it Matryoshka Skyrmions}, \href{https://doi.org/10.1016/j.nuclphysb.2013.03.003}{{\it Nucl.~Phys.} {\bf B 872} (2013) 62} [\href{https://arxiv.org/abs/1211.4916}{\tt arXiv:1211.4916}].

\bibitem{nitta4}
M.~Kobayashi and M.~Nitta, {\it Sine-Gordon kinks on a domain wall ring}, \href{https://doi.org/10.1103/PhysRevD.87.085003}{{\it Phys.~Rev.} {\bf D 87} (2013) 085003} [\href{https://arxiv.org/abs/1302.0989}{\tt arXiv:1302.0989}].

\bibitem{jennings}
P.~Jennings and P.~Sutcliffe, {\it The dynamics of domain wall Skyrmions}, \href{https://doi.org/10.1088/1751-8113/46/46/465401}{{\it J.~Phys.} {\bf A 46} (2013) 465401} [\href{https://arxiv.org/abs/1305.2869}{\tt arXiv:1305.2869}].

\bibitem{nitta5}
S.B.~Gudnason and M.~Nitta, {\it Domain wall Skyrmions}, \href{https://doi.org/10.1103/PhysRevD.89.085022}{{\it Phys.~Rev.} {\bf D 89} (2014) 085022} [\href{https://arxiv.org/abs/1403.1245}{\tt arXiv:1403.1245}].

\bibitem{GaLiRa}
V.A.~Gani, M.A.~Lizunova and R.V.~Radomskiy {\it Scalar triplet on a domain wall: an exact solution}, \href{https://doi.org/10.1007/JHEP04(2016)043}{{\it JHEP} {\bf 04} (2016) 043} [\href{https://arxiv.org/abs/1601.07954}{\tt arXiv:1601.07954}].

\bibitem{GaLiRaconf}
V.A.~Gani, M.A.~Lizunova and R.V.~Radomskiy, {\it Scalar triplet on a domain wall}, \href{https://doi.org/10.1088/1742-6596/675/1/012020}{{\it J.~Phys.: Conf.~Ser.} {\bf 675} (2016) 012020} [\href{https://arxiv.org/abs/1602.04446}{\tt arXiv:1602.04446}].

\bibitem{nitta6}
M.~Nitta, {\it Non-Abelian sine-Gordon solitons}, \href{https://doi.org/10.1016/j.nuclphysb.2015.04.006}{{\it Nucl.~Phys.} {\bf B 895} (2015) 288} [\href{https://arxiv.org/abs/1412.8276}{\tt arXiv:1412.8276}].

\bibitem{GaKiRu}
V.A.~Gani, A.A.~Kirillov and S.G.~Rubin, {\it Classical transitions with the topological number changing in the early Universe}, \href{https://arxiv.org/abs/1704.03688}{\tt arXiv:1704.03688}.

\bibitem{lensky}
V.A.~Lensky, V.A.~Gani and A.E.~Kudryavtsev, {\it Domain walls carrying a U(1) charge}, \href{https://doi.org/10.1134/1.1420436}{{\it Sov.~Phys.~JETP} {\bf 93} (2001) 677} [{\it Zh.~Eksp.~Teor.~Fiz.} {\bf 120} (2001) 778] [\href{https://arxiv.org/abs/hep-th/0104266}{\tt hep-th/0104266}].

\bibitem{GaKsKu01}
V.A.~Gani, V.G.~Ksenzov and A.E.~Kudryavtsev, {\it Example of a self-consistent solution for a fermion on domain wall}, \href{https://doi.org/10.1134/S1063778810110104}{{\it Phys.~Atom.~Nucl.} {\bf 73} (2010) 1889} [{\it Yad.~Fiz.} {\bf 73} (2010) 1940] [\href{https://arxiv.org/abs/1001.3305}{\tt arXiv:1001.3305}].

\bibitem{GaKsKu02}
V.A.~Gani, V.G.~Ksenzov and A.E.~Kudryavtsev, {\it Stable branches of a solution for a fermion on domain wall}, \href{https://doi.org/10.1134/S1063778811050085}{{\it Phys.~Atom.~Nucl.} {\bf 74} (2011) 771} [{\it Yad.~Fiz.} {\bf 74} (2011) 797] [\href{https://arxiv.org/abs/1009.4370}{\tt arXiv:1009.4370}].

\bibitem{Kudryavtsev1975}
A.E.~Kudryavtsev, {\it Solitonlike solutions for a Higgs scalar field}, \href{http://www.jetpletters.ac.ru/ps/1522/article_23290.shtml}{{\it JETP Lett.} {\bf 22} (1975) 82} [\href{http://www.jetpletters.ac.ru/ps/528/article_8373.shtml}{{\it Pis'ma v ZhETF} {\bf 22} (1975) 178}].

\bibitem{Anninos1991}
P.~Anninos, S.~Oliveira and R.A.~Matzner, {\it Fractal structure in the scalar $\lambda(\varphi^2-1)^2$ theory}, \href{https://doi.org/10.1103/PhysRevD.44.1147} {{\it Phys.~Rev.} {\bf D 44} (1991) 1147}.

\bibitem{Goodman2007}
R.H.~Goodman and R.~Haberman, {\it Chaotic scattering and the n-bounce resonance in solitary-wave interactions}, \href{https://doi.org/10.1103/PhysRevLett.98.104103} {{\it Phys.~Rev.~Lett.} {\bf 98} (2007) 104103} [\href{https://arxiv.org/abs/nlin/0702048}{\tt nlin/0702048}].

\bibitem{Campbell1983}
D.K.~Campbell, J.F.~Schonfeld and C.A.~Wingate, {\it Resonance structure in kink-antikink interactions in $\varphi^4$ theory}, \href{https://doi.org/10.1016/0167-2789(83)90289-0} {{\it Physica} {\bf D 9} (1983) 1}.

\bibitem{Peyrard1983}
M.~Peyrard and D.K.~Campbell, {\it Kink-antikink interactions in a modified sine-Gordon model}, \href{https://doi.org/10.1016/0167-2789(83)90290-7} {{\it Physica} {\bf D 9} (1983) 33}.

\bibitem{Campbell1986}
D.K.~Campbell, {\it Solitary wave collisions revisited}, \href{https://doi.org/10.1016/0167-2789(86)90161-2} {{\it Physica} {\bf D 18} (1986) 47}.

\bibitem{dorey}
P.~Dorey, K.~Mersh, T.~Romanczukiewicz and Y.~Shnir, {\it Kink-Antikink Collisions in the $\phi^6$ Model}, \href{https://doi.org/10.1103/PhysRevLett.107.091602}{{\it Phys.~Rev.~Lett.} {\bf 107} (2011) 091602} [\href{https://arxiv.org/abs/1101.5951}{\tt arXiv:1101.5951}].

\bibitem{GaKuLi}
V.A.~Gani, A.E.~Kudryavtsev and M.A.~Lizunova, {\it Kink interactions in the (1+1)-dimensional $\varphi^6$ model}, \href{https://doi.org/10.1103/PhysRevD.89.125009}{{\it Phys.~Rev.} {\bf D 89} (2014) 125009} [\href{https://arxiv.org/abs/1402.5903}{\tt arXiv:1402.5903}].

\bibitem{GaKuPRE} V.A.~Gani and A.E.~Kudryavtsev, {\it Kink-antikink interactions in the double sine-Gordon equation and the problem of resonance frequencies}, \href{https://doi.org/10.1103/PhysRevE.60.3305}{{\it Phys.~Rev.} {\bf E 60} (1999) 3305} [\href{https://arxiv.org/abs/cond-mat/9809015}{\tt cond-mat/9809015}].

\bibitem{oliveira01}
T.S.~Mendon\c ca and H.P.~de Oliveira, {\it The collision of two-kinks defects}, \href{https://doi.org/10.1007/JHEP09(2015)120}{{\it JHEP} {\bf 09} (2015) 120} [\href{https://arxiv.org/abs/1502.03870}{\tt arXiv:1502.03870}].

\bibitem{oliveira02}
T.S.~Mendon\c ca and H.P.~de Oliveira, {\it A note about a new class of two-kinks}, \href{https://doi.org/10.1007/JHEP06(2015)133}{{\it JHEP} {\bf 06} (2015) 133} [\href{https://arxiv.org/abs/1504.07315}{\tt arXiv:1504.07315}].

\bibitem{krusch01}
S.W.~Goatham, L.E.~Mannering, R.~Hann and S.~Krusch, {\it Dynamics of Multi-kinks in the Presence of Wells and Barriers}, \href{https://doi.org/10.5506/APhysPolB.42.2087}{{\it Acta Phys.~Polon.} {\bf B 42} (2011) 2087} [\href{https://arxiv.org/abs/1007.2641}{\tt arXiv:1007.2641}].

\bibitem{saad01}
D.~Saadatmand, S.V.~Dmitriev, D.I.~Borisov and P.G.~Kevrekidis, {\it Interaction of sine-Gordon kinks and breathers with a parity-time-symmetric defect}, \href{https://doi.org/10.1103/PhysRevE.90.052902}{{\it Phys.~Rev.} {\bf E 90} (2014) 052902} [\href{https://arxiv.org/abs/1408.2358}{\tt arXiv:1408.2358}].

\bibitem{saad02}
D.~Saadatmand et al., {\it Effect of the $\phi^4$ kink's internal mode at scattering on a PT-symmetric defect}, \href{http://www.jetpletters.ac.ru/ps/2075/article_31231.shtml}{{\it Pisma Zh.~Eksp.~Teor.~Fiz.} {\bf 101} (2015) 550} [\href{https://doi.org/10.1134/S0021364015070140}{{\it JETP Lett.} {\bf 101} (2015) 497}].

\bibitem{saad03}
D.~Saadatmand et al., {\it Kink scattering from a parity-time-symmetric defect in the $\phi^4$ model}, \href{https://doi.org/10.1016/j.cnsns.2015.05.012}{{\it Commun.~Nonlinear Sci.~Numer.~Simulat.} {\bf 29} (2015) 267} [\href{https://arxiv.org/abs/1411.5857}{\tt arXiv:1411.5857}].

\bibitem{rad1}
S.V.~Dmitriev, Y.S.~Kivshar and T.~Shigenari, {\it Fractal structures and multiparticle effects in soliton scattering}, \href{https://doi.org/10.1103/PhysRevE.64.056613} {{\it Phys.~Rev.} {\bf E 64} (2001) 056613}.

\bibitem{rad2}
S.V.~Dmitriev, P.G.~Kevrekidis and Y.S.~Kivshar, {\it Radiationless energy exchange in three-soliton collisions}, \href{https://doi.org/10.1103/PhysRevE.78.046604}{{\it Phys.~Rev.} {\bf E 78} (2008) 046604} [\href{https://arxiv.org/abs/0806.1152}{\tt arXiv:0806.1152}].

\bibitem{saad.arXiv.2016.08}
A.~Askari, D.~Saadatmand, S.V.~Dmitriev and K.~Javidan, {\it High energy density spots and production of kink-antikink pairs in particle collisions}, \href{https://arxiv.org/abs/1608.01847}{\tt arXiv:1608.01847}.

\bibitem{Deformed}
F.C.~Simas, A.R.~Gomes, K.Z.~Nobrega and J.C.R.E.~Oliveira, {\it Suppression of two-bounce windows in kink-antikink collisions}, \href{https://doi.org/10.1007/JHEP09(2016)104}{{\it JHEP} {\bf 09} (2016) 104} [\href{https://arxiv.org/abs/1605.05344}{\tt arXiv:1605.05344}].

\bibitem{Ahlqvist:2014uha}
P.~Ahlqvist, K.~Eckerle and B.~Greene, {\it Kink Collisions in Curved Field Space}, \href{https://doi.org/10.1007/JHEP04(2015)059}{{\it JHEP} {\bf 04} (2015) 059} [\href{https://arxiv.org/abs/1411.4631}{\tt arXiv:1411.4631}].

\bibitem{Mohammadi}
M.~Mohammadi and N.~Riazi, {\it Bi-dimensional soliton-like solutions of the nonlinear complex sine-Gordon system}, \href{https://doi.org/10.1093/ptep/ptu002}{{\it Prog.~Theor.~Exp.~Phys.} (2014) 023A03}.

\bibitem{weigel01}
H.~Weigel, {\it Kink-Antikink Scattering in $\varphi^4$ and $\phi^6$ Models}, \href{https://doi.org/10.1088/1742-6596/482/1/012045}{{\it J.~Phys.: Conf.~Ser.} {\bf 482} (2014) 012045} [\href{https://arxiv.org/abs/1309.6607}{\tt arXiv:1309.6607}].

\bibitem{weigel02}
I.~Takyi and H.~Weigel, {\it Collective coordinates in one-dimensional soliton models revisited}, \href{https://doi.org/10.1103/PhysRevD.94.085008}{{\it Phys.~Rev.} {\bf D 94} (2016) 085008} [\href{https://arxiv.org/abs/1609.06833}{\tt arXiv:1609.06833}].

\bibitem{baron01}
H.E.~Baron, G.~Luchini and W.J.~Zakrzewski, {\it Collective coordinate approximation to the scattering of solitons in the (1+1) dimensional NLS model}, \href{https://doi.org/10.1088/1751-8113/47/26/265201}{{\it J.~Phys.~A: Math.~and Theor.} {\bf 47} (2014) 265201} [\href{https://arxiv.org/abs/1308.4072}{\tt arXiv:1308.4072}].

\bibitem{javidan}
K.~Javidan, {\it Collective coordinate variable for soliton-potential system in sine-Gordon model}, \href{https://doi.org/10.1063/1.3511337}{{\it J.~Math.~Phys.} {\bf 51} (2010) 112902} [\href{https://arxiv.org/abs/0910.3058}{\tt arXiv:0910.3058}].

\bibitem{christov01}
I.~Christov and C.I.~Christov, {\it Physical dynamics of quasi-particles in nonlinear wave equations}, \href{https://doi.org/10.1016/j.physleta.2007.08.038}{{\it Phys.~Lett.} {\bf A 372} (2008) 841} [\href{https://arxiv.org/abs/nlin/0612005}{\tt nlin/0612005}].

\bibitem{GaKu}
V.A.~Gani and A.E.~Kudryavtsev, {\it Collisions of domain walls in a supersymmetric model}, \href{https://doi.org/10.1134/1.1423755}{{\it Phys.~Atom.~Nucl.} {\bf 64} (2001) 2043} [{\it Yad.~Fiz.} {\bf 64} (2001) 2130] [\href{https://arxiv.org/abs/hep-th/9904209}{\tt hep-th/9904209}, \href{https://arxiv.org/abs/hep-th/9912211}{\tt hep-th/9912211}].

\bibitem{manton_npb}
N.S.~Manton, {\it An effective Lagrangian for solitons}, \href{https://doi.org/10.1016/0550-3213(79)90309-2}{{\it Nucl.~Phys.} {\bf B 150} (1979) 397}.

\bibitem{kks04}
P.G.~Kevrekidis, A.~Khare and A.~Saxena, {\it Solitary wave interactions in dispersive equations using Manton's approach}, \href{https://doi.org/10.1103/PhysRevE.70.057603}{{\it Phys.~Rev.} {\bf E 70} (2004) 057603} [\href{https://arxiv.org/abs/nlin/0410045}{\tt arXiv:nlin/0410045}].

\bibitem{Radomskiy}
R.V.~Radomskiy, E.V.~Mrozovskaya, V.A.~Gani and I.C.~Christov, {\it Topological defects with power-law tails}, \href{https://doi.org/10.1088/1742-6596/798/1/012087}{{\it J.~Phys.: Conf.~Ser.} {\bf 798} (2017) 012087} [\href{https://arxiv.org/abs/1611.05634}{\tt arXiv:1611.05634}].

\bibitem{Aliakbar}
A.~Moradi Marjaneh, D.~Saadatmand, Kun Zhou, S.V.~Dmitriev and M.E.~Zomorrodian, {\it High energy density in the collision of $N$ kinks in the $\phi^4$ model}, \href{https://doi.org/10.1016/j.cnsns.2017.01.022}{{\it Commun.~Nonlinear Sci.~Numer.~Simulat.} {\bf 49} (2017) 30} [\href{https://arxiv.org/abs/1605.09767}{\tt arXiv:1605.09767}].

\bibitem{saad.prd.2015}
D.~Saadatmand, S.V.~Dmitriev and P.G.~Kevrekidis, {\it High energy density in multi-soliton collisions}, \href{https://doi.org/10.1103/PhysRevD.92.056005}{{\it Phys.~Rev.} {\bf D 92} (2015) 056005} [\href{https://arxiv.org/abs/1506.01389}{\tt arXiv:1506.01389}].

\bibitem{lohe}
M.A.~Lohe, {\it Soliton structures in $P(\varphi)_2$}, \href{https://doi.org/10.1103/PhysRevD.20.3120}{{\it Phys.~Rev.} {\bf D 20} (1979) 3120}.

\bibitem{khare}
A.~Khare, I.C.~Christov and A.~Saxena, {\it Successive phase transitions and kink solutions in $\phi^8$, $\phi^{10}$, and $\phi^{12}$ field theories}, \href{https://doi.org/10.1103/PhysRevE.90.023208}{{\it Phys.~Rev.} {\bf E 90} (2014) 023208} [\href{https://arxiv.org/abs/1402.6766}{\tt arXiv:1402.6766}].

\bibitem{GaLeLi}
V.A.~Gani, V.~Lensky and M.A.~Lizunova, {\it Kink excitation spectra in the (1+1)-dimensional $\varphi^8$ model}, \href{https://doi.org/10.1007/JHEP08(2015)147}{{\it JHEP} {\bf 08} (2015) 147} [\href{https://arxiv.org/abs/1506.02313}{\tt arXiv:1506.02313}].

\bibitem{GaLeLiconf}
V.A.~Gani, V.~Lensky, M.A.~Lizunova and E.V.~Mrozovskaya, {\it Excitation spectra of solitary waves in scalar field models with polynomial self-interaction}, \href{https://doi.org/10.1088/1742-6596/675/1/012019}{{\it J.~Phys.: Conf.~Ser.} {\bf 675} (2016) 012019}  [\href{https://arxiv.org/abs/1602.02636}{\tt arXiv:1602.02636}].

\end{thebibliography}
\end{document}